\documentclass[aps,pre,preprint,superscriptaddress,showpacs,floatfix]{revtex4-2}

\usepackage{graphicx}
\usepackage{amsmath,amssymb}
\usepackage{xcolor}
\usepackage{hyperref}
\usepackage{float}
\usepackage{subcaption}
\usepackage{multirow}
\usepackage{booktabs}
\usepackage{enumitem}
\usepackage{scalefnt}

\hypersetup{
    colorlinks=true,
    linkcolor=blue,
    citecolor=blue,
    urlcolor=blue,
}

\newcommand{\velvec}{\textbf{u}}

\begin{document}

\title{Effects of Geometric Modelling and Blood Rheology in Patient-Specific Arterial Blood Flow Simulations with Speed-Accuracy Trade-Off Analysis}

\author{Rishi Kumar}
\affiliation{Department of Chemical Engineering, Indian Institute of Technology, Kanpur, India 208016}

\author{K. Muralidhar}
\email{kmurli@iitk.ac.in}
\affiliation{Department of Mechanical Engineering, Indian Institute of Technology, Kanpur, India 208016}

\author{Indranil Saha Dalal}
\email{indrasd@iitk.ac.in}
\affiliation{Department of Chemical Engineering, Indian Institute of Technology, Kanpur, India 208016}
\begin{abstract}
This study investigates the effects of geometry modeling and reduction on predictions of blood flow in patient-specific descending aorta obtained from Computed Tomography (CT) scan data, followed by speed-accuracy trade-off analysis using 3D simulations. Additionally, we show how wall shear stresses (WSS) can be estimated for such realistic arteries using cheaper simulations of highly idealized equivalent geometries. The study uses realistic inflow (with mean Re = 313.6 and peak Re = 1687 based on Newtonian viscosity) and pressure outlet boundary conditions with Newtonian and non-Newtonian blood viscosity models, including the one developed by Apostolidis and Beris. Similar studies exist to predict the progression of atherosclerosis, clots, aneurysms and thrombus formation in arteries. However, these assume idealized flow geometries. In contrast, real arteries possess significant amounts of roughness, perturbations, and undulations. To understand their impact along with blood rheology, studies are performed with two levels of geometry reduction. Our results show the effects of such simplifications on flow field predictions in conjunction with blood rheology models. The first level of reduction loses the local asymmetry but is able to reasonably approximate various parameters, capturing patterns with the correct magnitude while showing significant computational speed-up. However, further simplification to an idealized smooth geometry loses all information about the vortex structures in the flow field. The flow recirculation zones near wall perturbations progressively weaken as the geometry is smoothed. Broadly, for blood rheology models, no significant differences are observed between the various types used in this study. Interestingly, for all models, the idealized smooth geometry, combined with area correction, yields WSS estimates that closely approximate those of the actual artery. In this study, we also present a detailed speed-accuracy trade-off analysis useful for future applications. The number of iterations to convergence becomes progressively lower with geometry reduction, with nearly four times speed-up obtained with the first level of simplification. Further, the computational time for the Newtonian model remains approximately an order of magnitude lower than the non-Newtonian counterparts.
\end{abstract}
\keywords{blood flow, descending aorta, geometry perturbations, pulsatile waveform, wall shear stress}
\maketitle

\section{Introduction}

Cardiovascular diseases (CVD) remain one primary reason for human mortality \cite{chatzizisis2007role} and have grown in importance after the COVID-19 pandemic. The progression of CVD is directly related to the geometric details of the arteries \cite{who_cvd_2017}. CVD is known to aggravate other conditions as well \cite{zheng2020covid}. Over the last few decades, increased computing power has enabled detailed computational studies of blood flow in human arteries \cite{javadzadegan2017haemodynamic,chaichana2011computation}. Researchers have been interested in the link between atherosclerosis and fluid flow and related parameters, including the wall shear stress (WSS) \cite{asakura_flow_1990,malek_hemodynamic_1999,ku_pulsatile_1985,stone_effect_2003}. Increasingly, studies have used patient-specific geometries intending to contribute to medical applications \cite{taylor_patient_specific_2009}. Similar studies have been reported to predict the progression of atherosclerosis \cite{lee_prediction_2019,knight_choosing_2010}, clots \cite{yesudasan_recent_2019}, aortic aneurysm\cite{peng2024preliminary}, and formation of thrombus \cite{cito_review_2013}. Significant studies have investigated the effect of stenosis on blood flow \cite{eshtehardi_association_2012,gholipour_three_dimensional_2018,karimi_finite_2013,mendieta2020importance}. However, in all these studies, the flow geometries have been assumed to be quite regular. In contrast, human arteries are far from regular and exhibit a significant amount of diameter variation, as well as variations in the direction of flow. The geometrical variations of arteries have been linked to the onset of atherosclerosis \cite{niu_surface_2013,schmidt_trucksass_quantitative_2003,cinthio_initial_2011}, thereby underscoring their importance in the broader context of human health. \textcolor{red}{In terms of hydrology, the work by Wood \cite{wood} discusses the averaging and scaling laws, which minimize the geometrical complexity of pores, thereby reducing the dimensionality of the problem.}

Despite the importance of irregularities in geometry, researchers have rarely focused on their impact on blood flow. Yi et al. \cite{yi2022impact} conducted simulations to investigate the effect of surface roughness on arterial stenosis. Shukla et al.\cite{shukla2020effect} studied the impact of surface roughness on pulsatile flow in a tube with a periodic pattern of roughness in the form of sinusoids, while Owen et al. \cite{owen2020assessment} studied the effect of surface roughness on blood flow in a tube. To the best of our knowledge, these three studies are the closest to examining the impact of roughness from a biological perspective in blood flow analysis. Among them, the studies by Yi et al. \cite{yi2022impact} and Shukla et al. \cite{shukla2020effect} use Newtonian fluid models. Owen et al. \cite{owen2020assessment} employed a range of models, including both Newtonian and non-Newtonian ones, to investigate the impact of an uneven geometry. However, the geometry and setup they created in the study can be argued to be far removed from reality. In a tube, they made a "rough" patch with surface irregularities that extended along its length and compared the fluid flow results between the rough patch and an equivalent smooth patch on the other side. This geometrical setup has two significant problems. Firstly, the roughness is distributed uniformly throughout the surface, and a tube with a rough and smooth patch is not closely aligned with the surface features of arteries in our body. Secondly, the surface features comprise not just roughness but also changes on the scale of the artery diameter. Hence, irregularities also exist in the shape and size fluctuations along the axial length of the arterial tube. The flowing fluid experiences the combined effect of these irregularities. As discussed later, the real artery may deviate significantly from the circular shape over a given plane. Additionally, the equivalent diameter can also show variations along the axis. To the best of our knowledge, no previous study has investigated the effect of such shape and size perturbations, in addition to surface roughness, which is the focus of the present study. Additionally, we employ a range of viscosity models to investigate the sensitivity of the non-Newtonian nature of blood to geometric perturbations.

Thus, in this study, we start with a patient-specific geometry – a part of the descending aorta with all possible details and perturbations retained for blood flow calculations. To simulate real-life conditions, we employ a range of viscosity models and realistic pulsatile inflow and pressure outlet boundary conditions. Recent developments in computational fluid dynamics have enabled the study of blood flow in realistic arterial geometries. Thus, it is possible to compute flow distribution in three dimensions for pulsatile flow in branches and bifurcations with bulges and constrictions. When the geometry is made realistic, undulations and perturbations of the artery wall need refinement in mesh size, thereby increasing computational cost. Secondly, for an accurate representation of a complex fluid such as blood, non-Newtonian rheology models must be used, further increasing the complexity. In the present study, we explore the consequences of simplifying the geometry in the context of various viscosity models (with an additional aim of reducing computational cost). The original artery model is obtained from CT scans, and progressively simpler geometries are generated by smoothing operations in image processing. We study hemodynamic aspects using Newtonian and non-Newtonian viscosity models in these equivalent geometries. Further, comparisons of results across these different geometries can yield insights into the importance of geometric resolution and perturbations on the flow field. 

To achieve this goal, three levels of equivalent geometries are examined. Level 1 is the original geometry derived from DICOM images provided by a medical hospital. Level 2 represents the same artery, divided into slices along its length, each slice representing a smooth duct of constant diameter. The number of slices can be varied to create varying levels of overall “smoothness”. Level 3 is a straight, smooth, circular tube with a linearly varying diameter, whose inlet and outlet diameters match the original. Inlet flow waveform is specified in terms of the peak flow rate and frequency. The significance of the study's outcome is manifold. The most direct one is linked to medical measurements and device perspectives. Alternatively, the study provides insight into the modeling aspects required for a given level of resolution of the selected artery. If we need to perform CFD calculations on an arterial geometry, the question of interest is, will a simplified geometry be sufficient? Or will the actual geometry with all small-scale details be needed to obtain medically relevant results? In addition, such studies also reveal the speed-accuracy trade-off. It is expected that the simplified geometric representation can be simulated in a lesser amount of computational time. However, some information is inevitably lost in the process of geometric simplification. For a particular application, one needs to anticipate the compromise in geometry permitted to perform a sufficiently accurate calculation in as little time as possible. This question is addressed in the present study by adopting equivalent geometries as described earlier. 

One additional aspect studied here is the change obtained by the choice of the viscosity model. Calculations in these three equivalent geometric representations provide intricate details of the effects of geometric perturbations on the flow of blood and their connection with the chosen rheological model, apart from the effect of these details on the computational speed. Additionally, we also ask an inverse question. Since highly idealized geometries are computationally cheaper, is it possible to use such simulations to obtain an estimate of relevant haemodynamic parameters in the patient-specific, real geometry with all perturbations and
undulations? In this study, we demonstrate the feasibility of this approach for estimating WSS, which is highly relevant in medical fields. The close agreement of such estimates and simulation results on the actual geometry, particularly for non-Newtonian models, opens up further
possibilities for future applications.

The outcomes of the simulations are presented in terms of vortical patterns and hemodynamic parameters, including the time-averaged wall shear stress (TAWSS) and oscillatory shear index (OSI). The accuracy losses are shown in terms of correlations defined in the article, which provides a numerical estimate of the loss of prediction accuracy across various simplification levels. The differences in flow patterns among the three levels are discussed in detail, with an understanding of the substantial reduction in CPU time. The results presented in this study may facilitate the simplification of complex arterial networks, thereby increasing the feasibility of meaningful full-body simulations in real-time. 
\section{Numerical Methodology}\subsection{Geometry} 
\begin{figure}[H]
    \centering
    \includegraphics[width=0.9\linewidth]{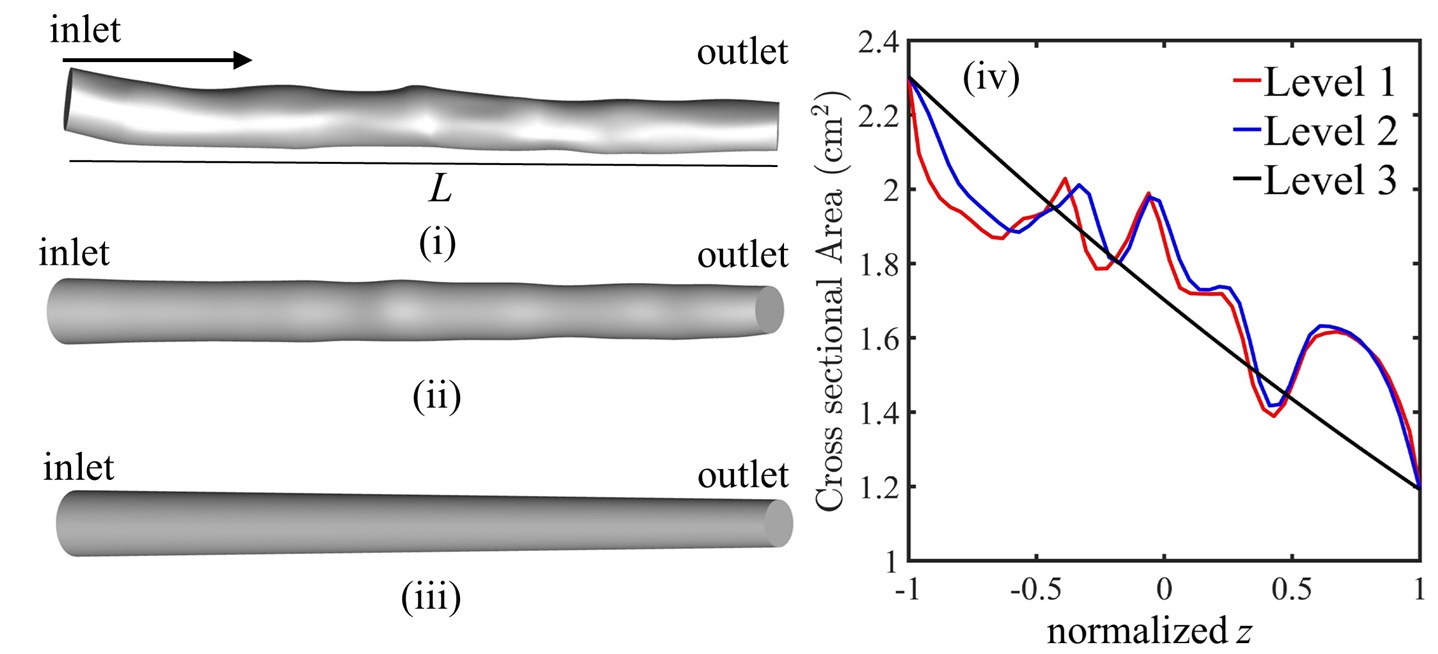}
    \caption{Level of geometrical aspects is shown, in (i) Level 1 of descending aorta extracted from CT-scan data, (ii) Slice-based geometry of descending aorta (Level 1 is divided into 25 slices for the construction of Level 2). (iii) Level 3 geometry based on the inlet and outlet diameter of Level 1 geometry. (iv) Cross-sectional area variation over the normalized z-axis for each level of simplification of the descending aorta.}
    \label{Refinement_of_geometry}
\end{figure}

The actual geometry of a descending aorta considered here is from patient-specific CT scan data obtained from the LPS Institute of Cardiology, Kanpur, India. The patient studied is a 40-year-old female, and her identity is otherwise anonymous. The reconstruction of the descending aorta is performed by the open-source software SimVascular v-2017 \cite{updegrove2017simvascular}, in which the pathline of the center of the artery is traced, followed by segmentation of the peripheral region of the artery and surface generation. The selected artery length and details of the cross-sectional areas are given in Table \ref{geometry_details}. For the present study, the descending aorta is simplified into three levels. Level 1 is the original patient-specific artery created by the DICOM images arising from a CT scan.\\ \textcolor{blue}{The simplification of the artery was performed based on the average area of the cross-section for every “slice”. We divided the Level 1 geometry into 25 equidistant slices (or segments) and measured the average cross-sectional area for each. This measured cross-sectional area is used to reconstruct the Level 2 geometry with circular cross-section throughout, keeping the area of cross-section equal for each slice/segment to the original geometry. Thus, for every such slice or segment, the average cross-sectional area remains the same and, most importantly, the asymmetry is also not retained. Thus, by increasing (or decreasing) the number of such segments, one can select the level of undulations to be preserved in Level 2. For Level 3, only the inlet and outlet cross-sectional areas are used, and a smooth, tapered geometry is created by joining both ends. After the reconstruction of the simplified level of the artery, a continuous measurement of cross-sectional area is taken and compared in Figure \ref{Refinement_of_geometry} (iv); the minute changes are due to the absence of the details between the slices in Level 2 only. However, it does not affect the major analysis of the simplification done in the study. }
 The three levels considered here represent a hierarchical simplification of the original geometry, retaining partial features of the original. It can be hypothesized that the grid requirements (and thus computational complexity) will be lower in smoother versions; however, the loss of accuracy regarding biomedical predictions remains to be investigated.

\begin{table}[ht]
\centering
\caption{\label{geometry_details}Geometrical details of the realistic artery (Level 1) considered in the study}
\begin{tabular}{lcr}
\toprule
Dimensions & Symbol & Value \\  
\midrule
Artery length & $L$ & 17.2 cm \\
Inlet cross-section area & $A_{\text{in}}$ & 2.3 cm$^{2}$\\
Outlet cross-section area & $A_{\text{out}}$ & 1.2 cm$^{2}$\\
\bottomrule
\end{tabular}
\end{table}

\subsection{Governing equations}
Patient-specific and simplified models of the arteries are used to study the hemodynamics of blood flow under pulsatile flow conditions. In this respect, the incompressible and unsteady forms of the continuity and momentum equations are solved in three dimensions with realistic boundary conditions and appropriate viscosity models.
The incompressibility constraint, in the form of the continuity equation and the applicable momentum equations for a variable viscosity fluid, is given below.
\begin{equation}
  \nabla \cdot \velvec = 0
\end{equation}
\begin{equation}
 \rho\left(\frac{\partial \velvec}{\partial t} + \velvec \cdot \nabla\velvec\right) = - \nabla p +  \nabla \cdot \Bar{\Bar\tau}
\end{equation}
Here $\velvec$ is the velocity vector, $p$  is pressure, and $\Bar{\Bar\tau}$ is the deviatoric part of the total stress tensor. The stress tensor is a  function of the strain rate given as 
$
\Bar{\Bar\tau} = \mu (\dot {\gamma})\Bar{\Bar\gamma}
$ , where $ \Bar{\Bar\gamma} = \frac{1}{2} ( \nabla \velvec +  \nabla \velvec^{T} )$. The magnitude of the strain rate is given as $ \dot \gamma = \sqrt{2\Bar{\Bar\gamma}:\Bar{\Bar\gamma}}$. Here $ \mu$ is the apparent viscosity of blood based on the selected viscosity model. \textcolor{red}{Here, body force (gravity) is neglected because its influence is negligible compared to that of pressure and inertia driving pulsatile flow in the descending aorta. The hydrostatic pressure variation along the modeled segment is small relative to the physiological pressure gradients. Therefore, including gravity would not significantly alter the predicted velocity and shear stresses \cite {shramko2023gravity}.}
\subsubsection{Boundary Conditions}
To solve the governing equations in the descending aorta and its simplified geometric versions, the following boundary conditions are applied:
\begin{figure}[ht]
    \centering
    \includegraphics[width=0.7\linewidth]{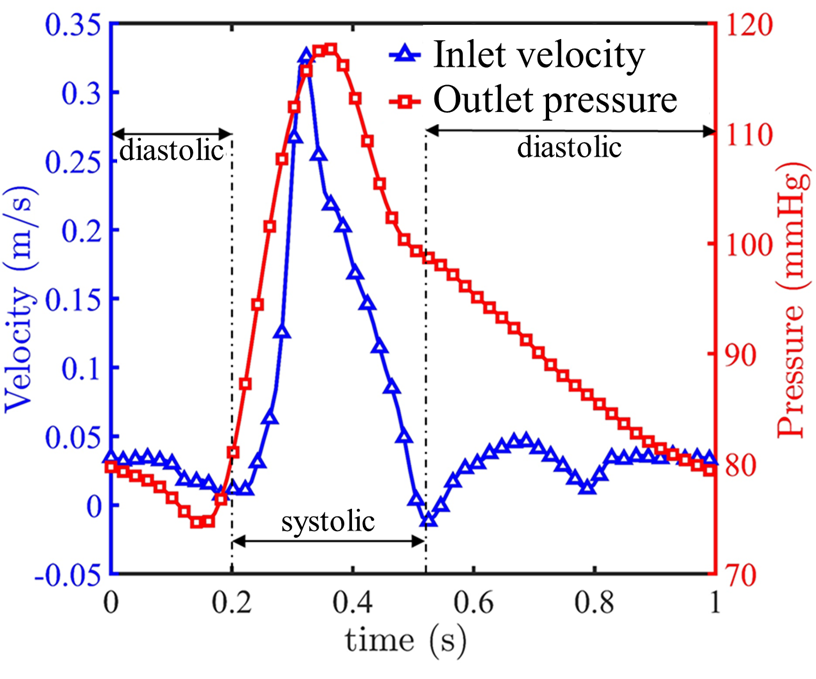}
    \caption{ Pulsatile velocity waveform incorporated at the inlet of the artery, along with the pulsatile pressure waveform at the outlet of the descending aorta. \cite{mills1970} . }
    \label{Inlet_outlet_profile}
\end{figure}
\begin{enumerate}
    \item {Inlet conditions:}  
    A uniform velocity profile is imposed at the inlets of all geometries, combined with a pulsatile waveform to replicate the periodic nature of blood flow under physiological conditions. This pulsatile waveform captures the unsteady characteristics of flow and ensures a realistic simulation of the hemodynamic environment. The pulsatile velocity profile is derived from physiological measurements reported by Mills et. al. \cite{mills1970}. \textcolor{red}{The inlet velocity and exit plane pressure profiles have been recorded simultaneously in time for a single patient and correspond to each other. This connectedness of velocity and pressure without a time lag is important.}\\ The derived profile is fitted by a Fourier series of up to 18 terms to apply the inlet boundary condition, with a mean Re of 313 and a peak Re of 1687. To fully allow the inflow to develop into a Womersley profile, the inlet region of the artery is extended by a length of \( 10D \), where \( D \) represents the characteristic diameter of the artery. This extension minimizes upstream flow irregularities resulting from flow development and ensures a realistic inflow condition within the computational domain of interest \cite{dong2024development}. Similarly, an extension of \( 5D \) is added to the outlet to mitigate numerical reflections of an outflow boundary condition and maintain a smooth flow transition at the exit plane.
    \item {Outlet conditions:}  
    Realistic pulsatile gauge pressure is applied at the outlets of the geometries, as depicted in Figure \ref{Inlet_outlet_profile}. These boundary conditions mirror the dynamic pressure variation observed in the cardiovascular system, which is critical for accurately capturing the pressure-driven nature of blood flow. The pulsatile pressure profile is derived from physiological measurements conducted by Mills et al. \cite{mills1970}, ensuring that it imposes real-world conditions. These outlet conditions play a crucial role in maintaining the accuracy of flow characteristics, particularly in scenarios where downstream vascular resistance and compliance significantly impact the upstream hemodynamics.
    \item {Wall conditions:}  
    The arterial walls in all geometries are assumed to be rigid, i.e., they do not deform under the influence of pressure and shear loading. A no-slip boundary condition is enforced at the walls, requiring the velocity at the wall to be zero. \textcolor{blue}{This assumption is widely used in computational blood flow simulations, as a rigid wall is expected to exhibit the strongest vortical structures.}
\end{enumerate}

\subsubsection{\large Blood viscosity models}
The selection of the viscosity model is crucial for simulating realistic blood flow characteristics. Blood is a non-Newtonian fluid, but since shear rates are smaller in vessels of large diameters, non-Newtonian effects are less pronounced here. For such arteries, earlier studies have shown only 
minor differences in predictions from Newtonian and non-Newtonian models in tubular geometries.
For the geometry considered in the present study, surface perturbations are present in Level 1, and differences in predictions across models may be expected. Thus, in this study, we consider the Newtonian and two non-Newtonian models for blood viscosity. The selected viscosity models are discussed below.
\begin{enumerate}
\item Newtonian\\
The Newtonian model \cite{merrill1967viscosity} is widely employed in blood flow simulations, primarily due to its linearity and the resulting computational efficiency. It is expected that for arteries of relatively larger size, Newtonian model predictions are negligibly
different from the non-Newtonian ones. In the present study, blood viscosity is taken to be a constant, namely
\begin{equation}
     \mu_{app} = 0.0035 \;\text{Pa s}
\end{equation}
\noindent It is appropriate for a body temperature of 310 K.

\item Carreau-Yasuda\\
The Carreau-Yasuda \cite{cho1991effects} is one of the most widely used Generalized Newtonian fluid (GNF) models. For viscoelastic fluids, the Carreau-Yasuda model is more realistic than the Power-law model, as it exhibits a plateau at both low and high shear rates. Specifically, blood demonstrates near-constant viscosity at high shear rates, consistent with the assumptions of a Newtonian fluid under such conditions. The Carreau-Yasuda viscosity model is given as:
\begin{equation}
  \mu_{app} =\mu_{0} + (  \mu_{0} -  \mu_{\infty} ) (1 + (\lambda \dot \gamma)^{P})^{\frac{n-1}{P}}
\end{equation}
In this context, $\mu_{0}$ and $\mu_{\infty}$ are the saturation values of viscosity at low and high shear rates, respectively. The parameter $P$ is known as the Yasuda exponent, with both $P$ and $n$ determining the degree of nonlinearity. The time constant $\lambda$ regulates the range of strain rates at which the transition to a power-law type regime is obtained. In this way, $\lambda$ is related to the relaxation time of the fluid. The model simplifies to the Newtonian fluid when the time constant is zero ($\lambda$ = 0). Table \ref{CY_viscosity_parameter} lists the parameters used for the Carreau-Yasuda model for blood in this study.

\begin{table}[ht]
\centering
\caption{\label{CY_viscosity_parameter}Parameters considered for the Carreau-Yasuda blood viscosity model at a body temperature of 310 K}
\begin{tabular}{l c} 
\toprule
Carreau-Yasuda Viscosity Parameters & Value \\  
\midrule
$ \mu_{0} $ & 0.056 Pa·s \\ 
$ \mu_{\infty} $ & 0.0035 Pa·s \\ 
$ \lambda $ & 3.313 s \\ 
$ n $ & 0.3568 \\ 
$ P $ & 2 \\ 
\bottomrule
\end{tabular}
\end{table}

\item Apostolides and Beris \\
The recently developed viscosity model by Apostolides and Beris \cite{apostolidis2014modeling} is based on the Casson viscosity model. The parameters are functions of the constituents of blood, such as hematocrit concentration $\phi$ and fibrinogen c$_{f}$, whose values are
adjusted using various flow-related data available in the literature. The apparent fluid viscosity is given by the following equation:
\begin{equation}
     \mu_{app} = \left(\sqrt{\frac{\tau _{o}}{\dot \gamma}} + \sqrt{\mu _{c}}\right)^2
\end{equation}
 Here, $\dot \gamma$ is the magnitude of the strain rate while the yield stress $\tau_{0}$ is given as follows:
\begin{equation}
    \tau_{0} = \begin{cases}
    (\phi\ - \phi_c)^{2}[0.508c_f + 0.4517]^{2}, & \text{if $\phi > \phi_c$};\\
    0, & \text{if $\phi  \leq \phi_c$};
\end{cases}
\end{equation}
A threshold value of hematocrit is required to calculate the yield stress; this threshold is a quadratic function of the fibrinogen concentration.
\begin{equation}
     \phi_{c} = \begin{cases}
    0.312c_{f}^{2} -0.468c_{f} +0.1764, & \text{ $ c_{f} < 0.75$};\\
    0.0012; & \text{ $ c_{f} \geq 0.75$};
\end{cases} 
\end{equation}
\begin{figure}[H]
    \centering
    \includegraphics[width=0.6\linewidth]{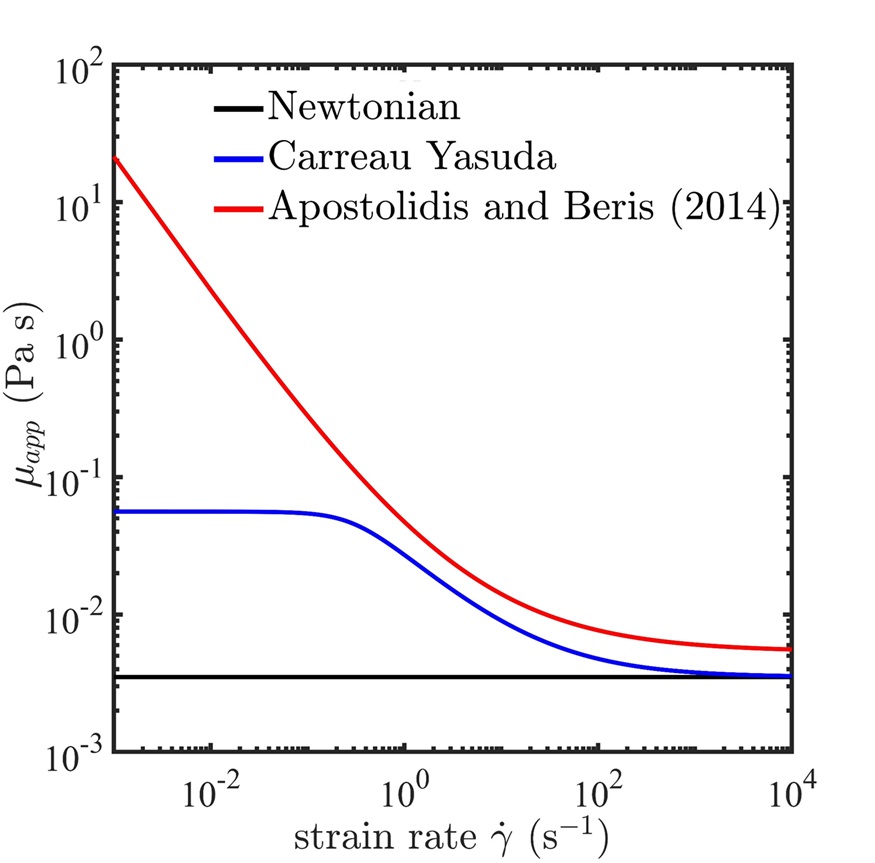}
    \caption{Blood viscosity models selected for simulating the blood flow in the descending aorta. Newtonian, Carreau Yasuda, and Apostolides \& Beris models are shown.}
    \label{Viscosity}
\end{figure}
The Casson viscosity, i.e., $\mu_c$, is also obtained by using the experimental data available in the literature and follows an Arrhenius dependence on temperature, given as:
\begin{equation}
    \mu_{c} = \mu_{p}(1+2.0703\phi +3.722\phi^{2})\exp \left[-7.0276\left(1 - \frac{T_{0}}{T} \right)\right] 
\end{equation}

At fully developed conditions, the local shear rate is zero at the center of a smooth circular channel. This local shear rate will make the apparent viscosity infinitely large, leading to an instability in the numerical solver.  To ensure the stability of the solver at fully developed conditions, Miller and Morris \cite{miller2006normal} suggested that the shear rate magnitude is regularised as $\dot \gamma = \dot \gamma + \dot \gamma_{r} $, where the non-local shear rate $\dot \gamma_{r} << \dot \gamma$ and given as 
$$
\dot \gamma_{r} = \frac{\epsilon U_{max}}{L_c}
$$
Here, $U_{max}$ is the maximum velocity attained in the Newtonian viscosity model, $\epsilon=0.01$ is the regularization parameter, and $L_{c}$ is the characteristic length. In the present study, it is chosen as the inlet diameter of the Level 1 artery is chosen. The parameters selected are appropriate for a normal human body temperature of 310 K and are summarized in Table \ref{AB_viscosity_parameter}.

\begin{table}[ht]
\centering
\caption{\label{AB_viscosity_parameter}Parameters considered for the Apostolidis and Beris blood viscosity model\cite{apostolidis2014modeling}}
\begin{tabular}{l c} 
\toprule
Blood Parameters for the Apostolidis and Beris Viscosity Model & Value \\   
\midrule
Average hematocrit concentration, $ \phi$ & 0.4  \\ 
Temperature of Blood Sample, $ T$ & 310 K  \\ 
Reference Temperature, $T_{0}$ & 296.1 K \\ 
Plasma viscosity, $\mu_{c}$ & $1.67 \times 10^{-3}$ N·s/m$^{2}$ \\ 
Density of blood, $\rho$ & 1060 kg/m$^{3}$ \\ 
Fibrinogen concentration, $ c_{f}$ & 0.125 g/dl \\ 
\bottomrule
\end{tabular}
\end{table}

\end{enumerate}
\par

\section{Flow distribution in a descending aorta}
\subsection{Grid Independence Test and Validation}
\begin{figure}[H]
    \centering
    \includegraphics[width=\linewidth]{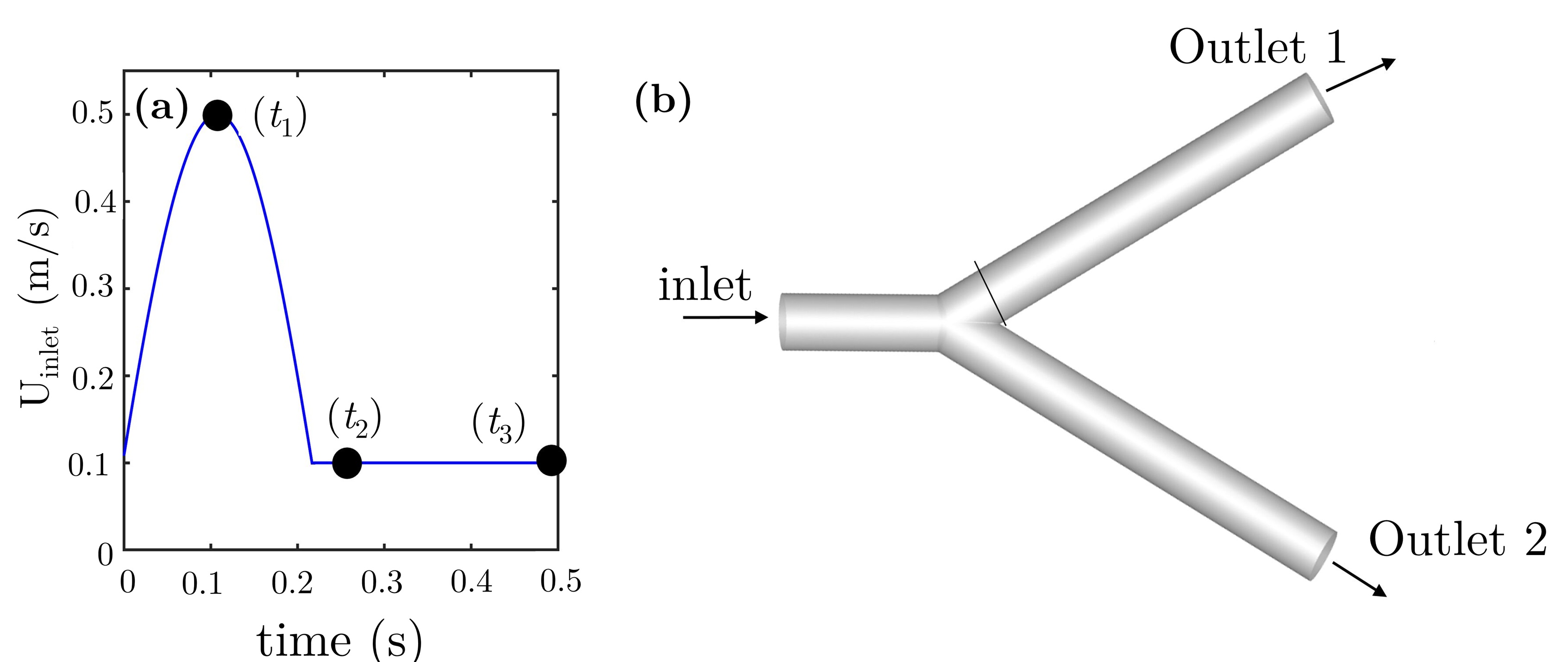}
  \caption{(a) Oscillatory velocity inlet profile. (b) symmetric bifurcation geometry used by  Nagargoje et al. \cite{nagargoje2021pulsatile}}
    \label{validation_figure}
\end{figure}

\begin{figure}[H]
    \centering
    \includegraphics[width=\linewidth]{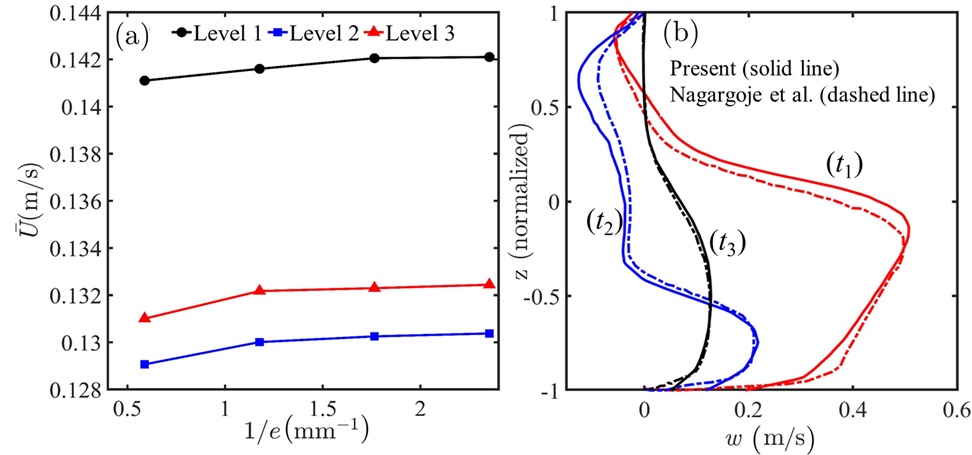}
  \caption{(a) Mesh independence test is performed for Levels 1-3 of the descending aorta. The average velocity at the outlet is presented as a function of the inverse of the mesh element size. Here, $e$ is the mesh element size, and a higher inverse element size indicates a finer mesh. (b) The \( z \)-velocity component (\textit{w}) is plotted at time instants \( t_{1}, t_{2}, \) and \( t_{3} \) at the bifurcation of the geometry and matched against comparable results of Nagargoje et al. \cite{nagargoje2021pulsatile}}
    \label{GIT_validation}
\end{figure}

3D simulations of the flow distribution in the descending aorta in patient-specific (Level 1) and its simplified versions (Levels 2 and 3) are performed using the finite volume method within ANSYS-Fluent\textregistered, a commercial flow solver. For the grid independence test for the three geometries, the steady flow problem is solved on four refined meshes (M1, M2, M3, and M4). The solver used in this study is also validated with the results of Nagargoje et al. \cite{nagargoje2021pulsatile} for a symmetric bifurcation with pulsatile inlet boundary conditions. The pulsatile waveform and the geometry of interest are shown in Figure \ref{validation_figure}.\\ \textcolor{blue}{A steady flow problem is used for the grid independence test, instead of a pulsatile flow input. This is performed as a measure to save computational time and fix mesh sizes at the start, since pulsatile flow simulations incur much higher costs. The steady uniform inlet velocity (which itself is the average of the pulsatile waveform) is given to the physical domain of all arteries. Since the goal was to find the spatially converged grid, a single parameter is used to compare the results. The outlet average velocity is the simplest yet effective parameter used in various studies. In CFD simulations of incompressible flow, the SIMPLE algorithm (or its variant) does not exactly satisfy the mass balance constraint (i.e., the equation $\nabla\cdot u = 0$ is not exactly solved). Hence, it is necessary to examine the divergence-free condition at the cell level (here 1E-5) and jointly check if the inflow and outflow values of average velocity match to within 10-3 \% at all times. The former constraint at the cell level is also implemented in pulsatile flow simulations.} In Figure \ref{GIT_validation}(a), the average velocity obtained is shown as a function of the inverse of the mesh element size. Following grid independence, the M3 mesh is selected for further simulations. Validation against the work of \cite{nagargoje2021pulsatile} is shown in Figure \ref{GIT_validation}(b). The $z$-component velocity is plotted on the centerline near the bifurcation for three time instants of the oscillatory inflow condition. The validation results are satisfactory, providing a check on the solver's correctness.
Next, flow metrics and hemodynamic parameters are analyzed to understand the implications of the geometric simplifications for the three viscosity models. Quantities of interest include the velocity profiles and wall shear stress, along with hemodynamic parameters such as the time-averaged wall shear stress (TAWSS) and the Oscillatory Shear Index (OSI). Analysis in terms of the Q-criterion for strain rate and rotation is also reported.
\subsection{Velocity Profiles}
\begin{figure}[H]
    \centering
    \includegraphics[angle=0,width=1.05\linewidth]{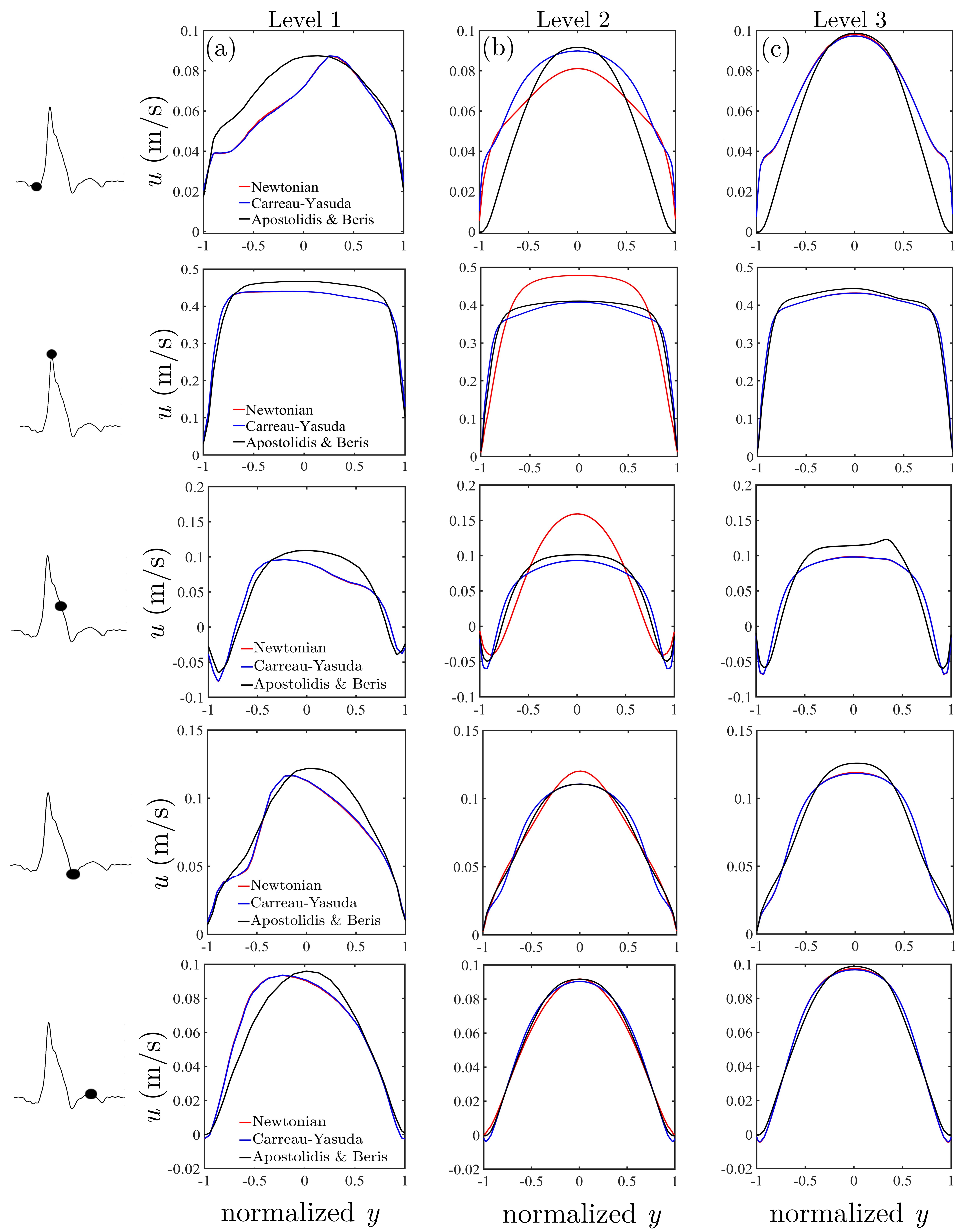}
  \caption{Axial velocity component shown for the three different viscosity models in (a) Level 1, (b) Level 2, and (c) Level 3 geometries.}
    \label{velocity_plots_levels}
\end{figure}

\begin{figure}[H]
    \centering
    \includegraphics[angle=0,width=1\linewidth]{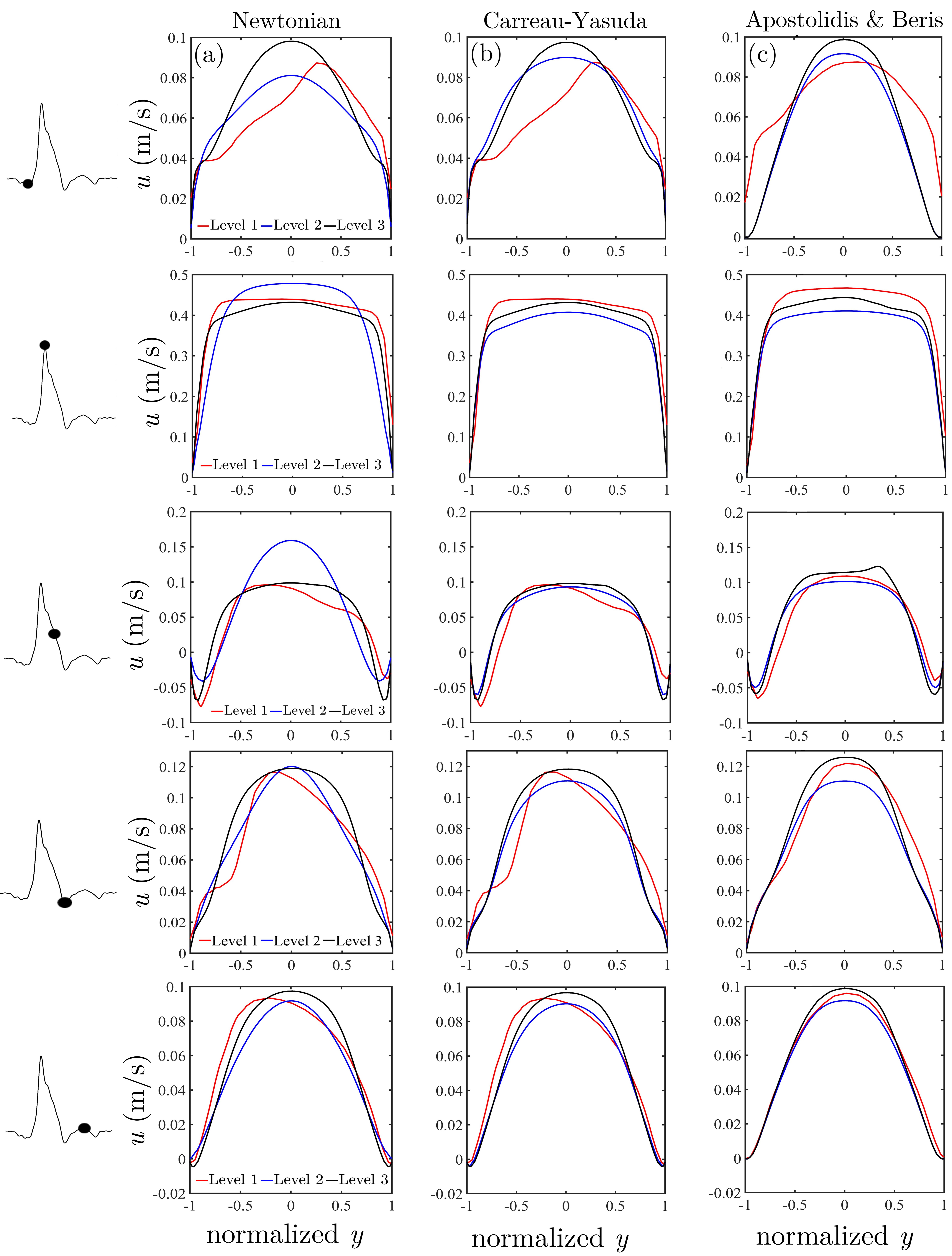}
  \caption{Axial velocity component shown for the original and two simplified geometries (Levels 1-3) for (a) Newtonian, (b) Carreau-Yasuda, and (c) Apostolidis and Beris viscosity models.}
    \label{velocity_plots_viscosity}
\end{figure}

This section presents the cross-sectional velocity distribution during the systolic and diastolic phases over various planes of the descending aorta for the three viscosity models and the Level 1-3 geometries. The velocity data are shown in Figures \ref{velocity_plots_levels} and \ref{velocity_plots_viscosity} for the 4th cycle of simulation, as the flow was observed to exhibit apparent periodicity after the third cycle, indicating a fully developed dynamic steady state. Specific time instants from the cycle, as highlighted in the left-most columns of Figures \ref{velocity_plots_levels} and \ref{velocity_plots_viscosity}, are used for discussion.

\begin{figure}[H]
    \centering
    \begin{subfigure}[b]{0.51\textwidth} 
        \centering
        \includegraphics[width=\textwidth]{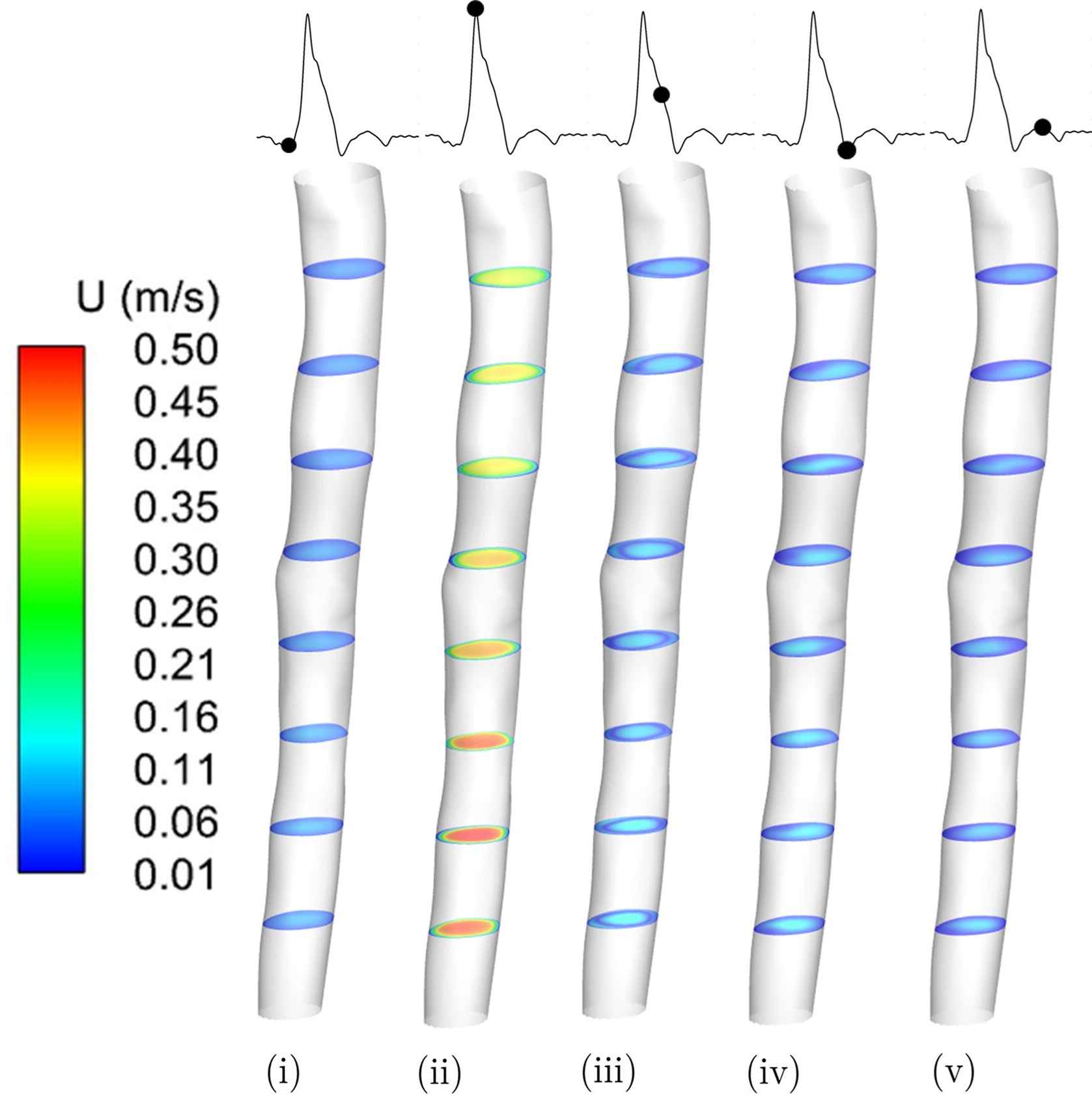}
        \caption{}
        \label{velocity_slice}
    \end{subfigure}
    \hfill
    \begin{subfigure}[b]{0.4\textwidth}
        \centering
        \includegraphics[width=\textwidth]{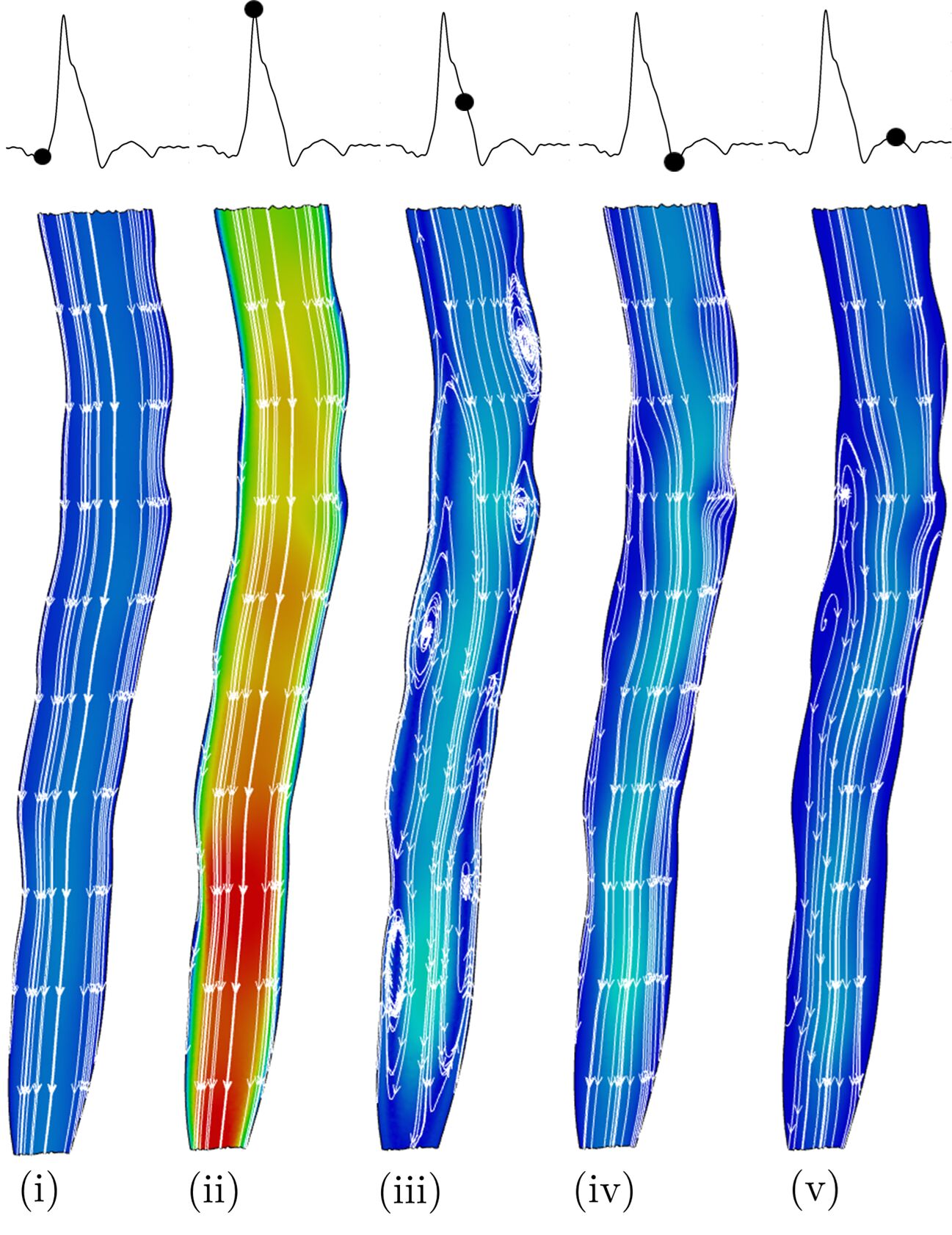}
        \caption{}
        \label{velocity_streamline}
    \end{subfigure}
    \caption{(a) Velocity magnitude contours at various time instants for Level 1 artery and (b) streamlines superimposed on the velocity magnitude contour over the median plane for the  Apostolidis and Beris viscosity model. (i)-(iv) show distinct phases of the vascular waveform.}
    \label{velocity_comparison}
\end{figure}

\begin{figure}[H]
    \centering
    \includegraphics[angle=0,width=1\linewidth]{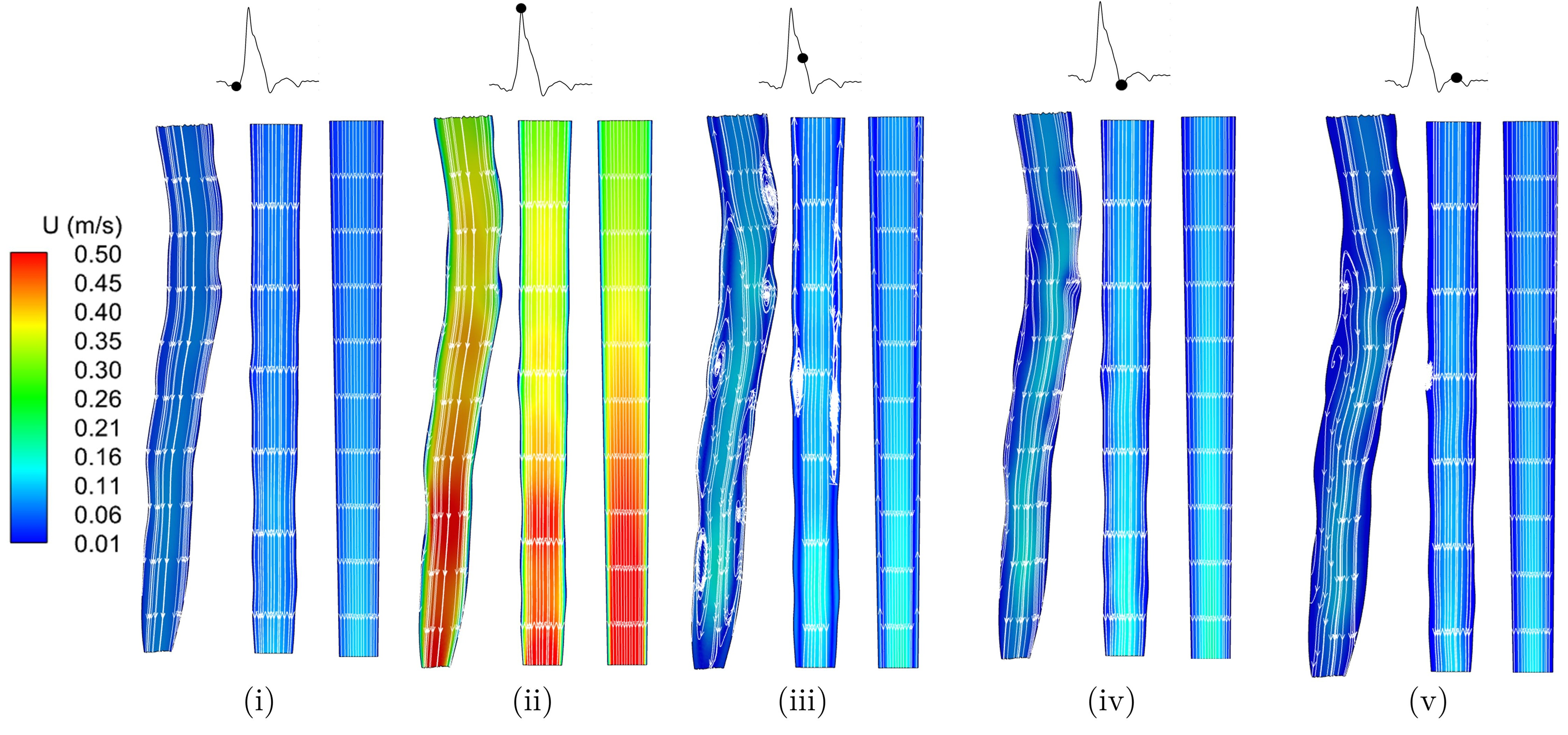}
  \caption{Velocity contours superimposed with streamlines are shown at selected time instants for the Apostolidis and Beris viscosity model for the three geometry levels 1-3. Within (i)-(v), Levels 1 to 3 are shown from left to right.}
    \label{streamline_level}
\end{figure}

\textcolor{blue}{The systolic phase of the waveform (Figure 2) includes three key instants: \( t_1 \) = 0.2 s in the initial instant of increase in inflow velocity, \( t_2 \) = 0.3 s is the peak (of accelerating phase), and \( t_3 \)= 0.45 s is the deceleration phase \cite{owais2023effect}. The diastolic phase is analyzed at two time instants, \(t_4 = 0.6 \) s and \(t_5 = 0.7 \) s, representing flow fluctuations towards the end of the cycle.}
This section further describes the velocity characteristics across the cross-section in the form of centerline plots, along with streamlines, contours, and vectors, for the selected viscosity models and geometric simplification levels.
In Figures \ref{velocity_plots_levels} and \ref{velocity_plots_viscosity}, the distribution of the axial velocity component (denoted by $u$) is shown at the centerline of the cross-section located at a distance of $z = 16D$ from the extended inlet as a function of the normalized $y$-coordinate. We present a comparison across two factors. In the first set of figures (Figure \ref{velocity_plots_levels}), we focus on comparing the viscosity models for a particular level of simplification, highlighting subtleties in model predictions. In Figure \ref{velocity_plots_viscosity}, we show a comparison across the level of simplification for any given viscosity model. This completes the picture, showing how modeling aspects are coupled with geometric details. In Figure \ref{velocity_plots_levels}, each level of geometry simplification is shown in a column, while rows represent distinct phases of the pulsatile flow cycle. The velocity profiles, normalized across the channel width, are depicted by red, blue, and black lines for the Newtonian, Carreau-Yasuda, and Apostolidis-Beris models, respectively. The axial velocity component distribution is shown at selected time instants for the individual viscosity models, corresponding to a particular level of geometry simplification. An asymmetric velocity distribution is observed for Level 1 only. This is expected since the perturbation and undulations across the circumference at any axial position are retained in Level 1, which leads to an asymmetric flow in succeeding cross-sections. Interestingly, in Level 1, the flow profiles are nearly identical for all phases of the cycle for the Newtonian and Carreau-Yasuda viscosity models. This similarity is expected since the Newtonian and Carreau-Yasuda models predict similar viscosities at high shear rates (Figure 3). The velocity predictions for the Apostolidis and Beris viscosity model are slightly larger than the other two for all the phases. For Level 2 (Figure \ref{velocity_plots_levels}(b)), the geometry is simplified and made symmetric around the circumference, and symmetric flow profiles are observed for all the viscosity models at every phase of the pulsatile flow. Interestingly, at this level, the two non-Newtonian models - Carreau-Yasuda and Apostolidis and Beris- predict similar velocity profiles, whereas the Newtonian predictions show an overshoot near the centerline. Note that such differences at the centerline are known between Newtonian and non-Newtonian models for flows in a smooth duct, where the latter needs a resolution of the shear rate singularity within the definition of apparent viscosity. For the simplest Level 3 in Figure \ref{velocity_plots_levels}(c), which is a tapered smooth tube, the predictions across viscosity models are largely symmetric and similar to each other, owing to the symmetry of the geometry.
Next, in Figure \ref{velocity_plots_viscosity}, we present a comparison of velocities across different geometry simplifications for each viscosity model. The data shows an accuracy check for such simplifications across different viscosity models considered in the present study. Columns (a), (b), and (c) depict the velocity profiles for the Newtonian, Carreau-Yasuda, and Apostolidis and Beris viscosity models, respectively. As mentioned earlier, the rows represent distinct phases of the pulsatile flow cycle. Essentially, the trends agree with each other, except for the asymmetry that is only visible for the patient-specific geometry (Level 1). The most significant differences are observed at the start of the cycle (first row) and during the deceleration phase (third row). Interestingly, for the deceleration phase, only the predictions with the combination of Level 3 and Newtonian fluid are significantly more than the others. Thus, with respect to velocity profiles, the simplified geometry can capture the trends quite well for all viscosity models, except for the expected asymmetry in Level 1. 
The following sections present further analyses, including streamlines, velocity contours, and vectors. Figure \ref{velocity_slice} shows velocity magnitude contours at various cross-sectional planes in the descending aorta in the Level 1 geometry for multiple phases of the inflow waveform. Additionally, Figure \ref{velocity_streamline} displays the velocity magnitude combined with streamline plots over the medial plane of the Level 1 aorta. These details are presented here for the Apostolidis and Beris viscosity model. The key trends remain similar to those of the other two models and are therefore not repeated. During the systolic phase, the inflow increases steadily, reaching a peak corresponding to the maximum flow rate. This progressive increase in inflow elevates wall forces, consequently enhancing wall loading. In contrast, the diastolic phase exhibits a much more complex flow behavior, driven by the adverse pressure gradient that develops as the inflow decelerates in time. The adverse gradient generates regions of backflow where the flow reverses direction near the walls.
These regions are particularly pronounced in areas with irregular geometry, abrupt changes in cross-section, and surface perturbations. The backflow generated disrupts the unidirectional flow and interacts with the surrounding fluid, leading to flow separation and the formation of recirculation zones. The interplay between backflow, flow circulation, and pressure gradients highlights the sensitivity of the flow field to boundary conditions and geometric variations, reinforcing the importance of retaining detailed geometry. Thus, a loss in prediction accuracy is expected with geometry simplifications, some of which are examined next in Figure \ref{streamline_level}. 
Streamlines in the three geometries at all time instants are shown in Figure \ref{streamline_level} with the Apostolidis and Beris viscosity model. It is observed that during the systolic phase, all the streamlines are unidirectional, and no flow circulation is observed. However, as the flow enters the diastolic regime, flow circulations appear in both Level 1 and Level 2, but are absent in Level 3. These flow recirculation zones result from the interplay of geometrical perturbations and adverse pressure gradient conditions. With the absence of perturbations in Level 3, such zones of backflow are absent. Reversed flow zones are also larger in Level 1, relative to Level 2, owing to geometric variations coupled with a loss of symmetry. Thus, as the artery geometry is simplified, the intensity of circulation decreases and disappears in the limit of the smooth geometry approximation, namely, Level 3. Flow circulation patterns are essential from a medical perspective, as they influence the wall shear stress (WSS) and pressure loading. A stagnant circulation region will promote particulate deposition, leading to thrombosis. Hence, delineating such areas is essential from a biomedical perspective. 
\textcolor{red}{The pressure variation along different arterial levels is illustrated in Figure \ref{Pressure_contours} for selected time instants of the cardiac cycle. The pressure values are presented relative to the physiological human blood-pressure range. It is observed that the geometric simplification of the descending aorta has a minimal influence on pressure, as the overall trend remains consistent at each time instant. However, the expected variation in pressure associated with changes in the inlet condition is clearly visible.}

\textcolor{red}{
\begin{figure}[H]
    \centering
    \includegraphics[angle=0,width=1\linewidth]{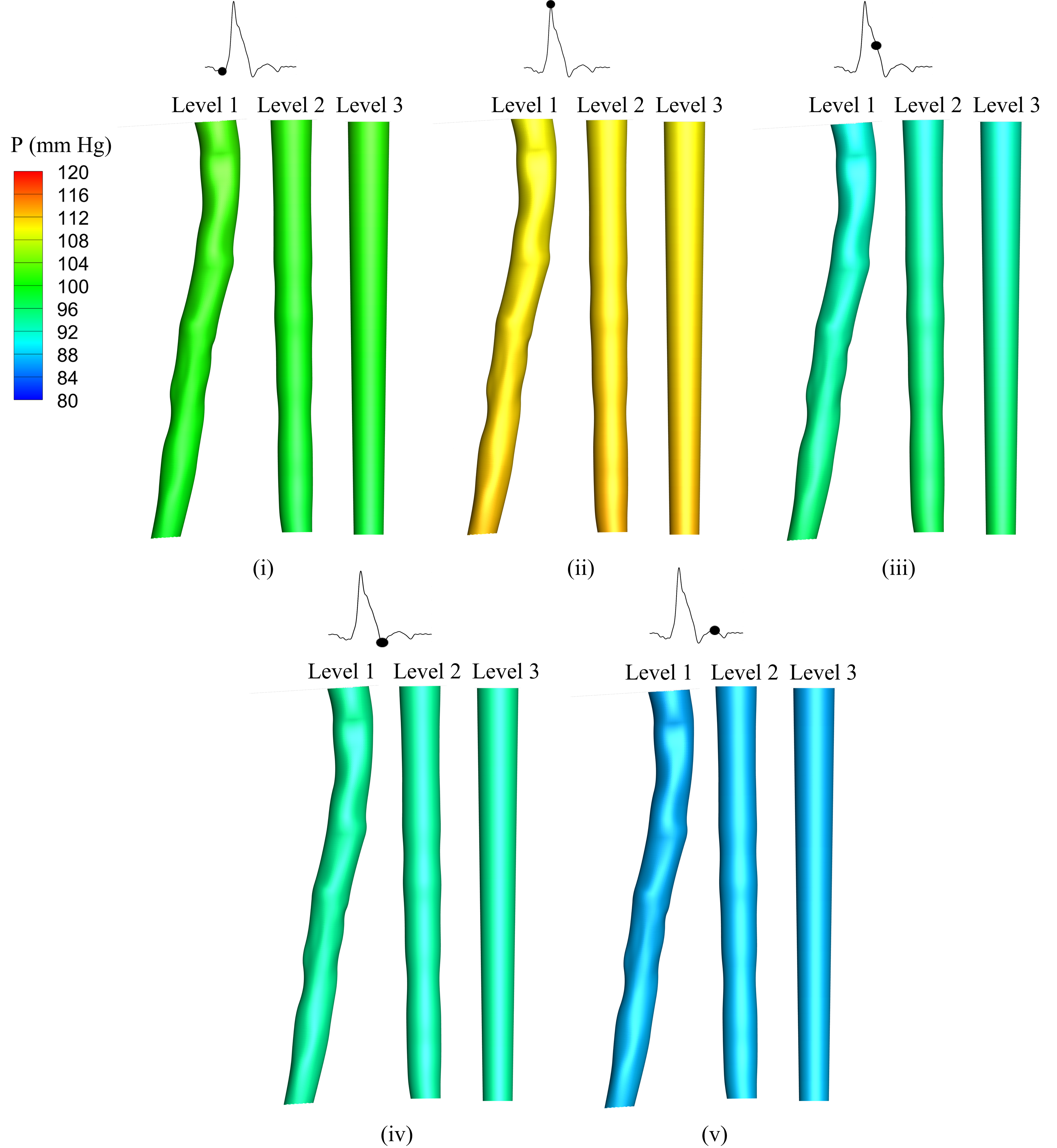}
  \caption{Pressure contour shown for the simplified geometries for the Apostolidis and Beris viscosity model at the selected time instant.}
    \label{Pressure_contours}
\end{figure}
}
\subsection{Hemodynamic Parameters}
\subsubsection{\large  Wall shear stress (WSS)}
\begin{figure}[H]
    \centering
    \includegraphics[width=\linewidth]{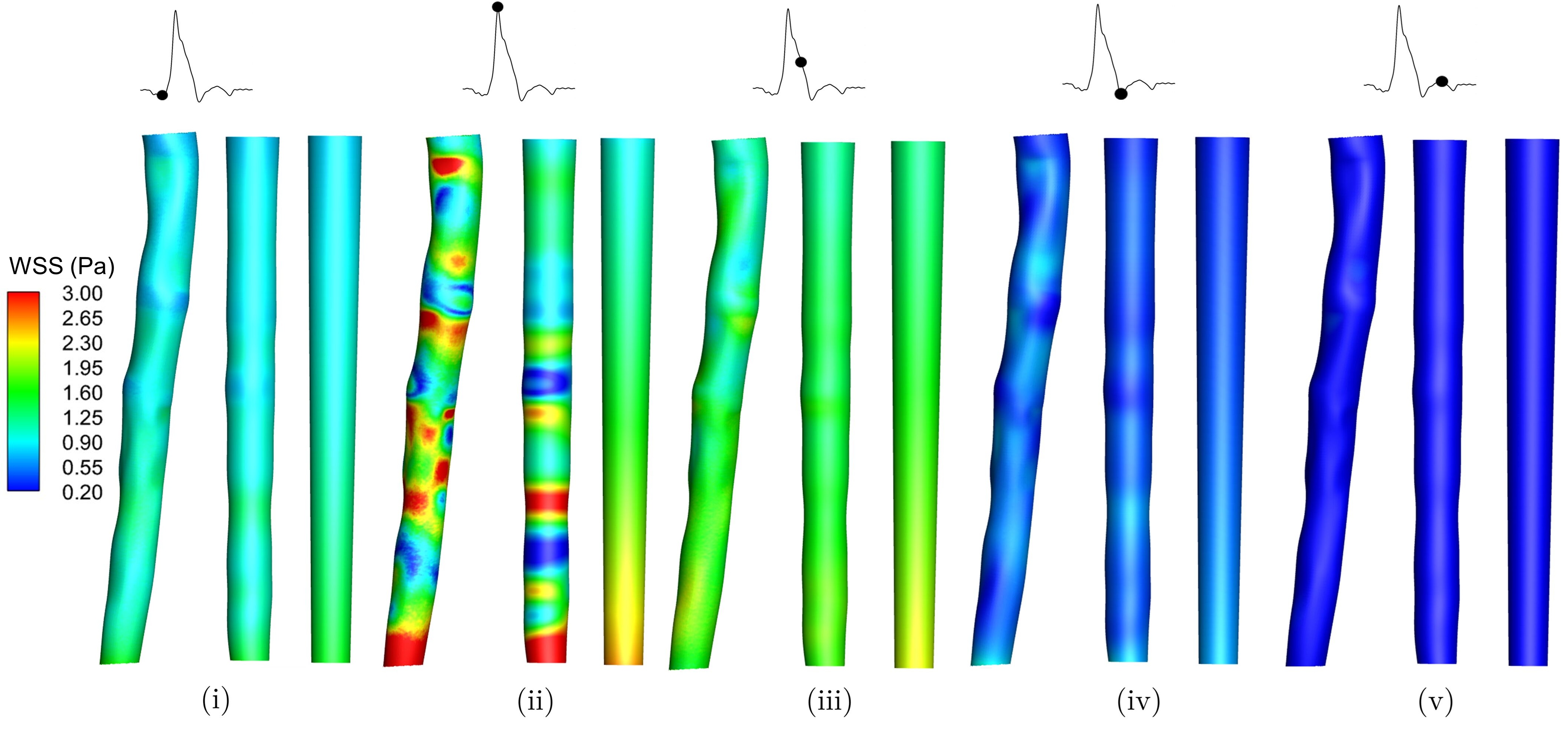}
  \caption{Wall shear stress (WSS) contours superimposed with streamlines are shown at the selected time instants of the vascular cycle for the Apostolidis and Beris viscosity model. Within (i) to (v), Levels 1 to 3 are shown from the left.}
    \label{WSS}
\end{figure}

Wall shear stress (WSS) plays a crucial role in regulating the endothelial function and has been widely recognized as a key factor in the development of atherosclerosis. Endothelial cells sense the wall shear stress via multiple responsive cell components, and activation or deactivation of intracellular pathways leads to gene and protein expression regulation\cite{gijsen2019expert}. This continuous process is essential for remodeling, development, and growth of the vascular walls. Moreover, the clinical significance of wall shear stress lies in its ability to predict the risk of future cardiovascular events, making it an important biomarker in cardiovascular health.
Mathematically, the shear stress can be calculated with the help of velocity gradients
$( \nabla \velvec)$ as follows:
\textcolor{red}{
\begin{equation}
\Bar{\Bar\tau}_{} = 2\mu_{app}\Bar {\Bar {\gamma}}
\end{equation}
The strain rate ($ \Bar{ \Bar{ \gamma}}$) is given by the sum of the velocity gradient and the transpose of the velocity gradient $\Bar {\Bar {\gamma}} = ( \nabla \textbf{u} +  \nabla \textbf{u}^{T})/2$ at the wall. \\For the wall shear stress, the traction vector of the shear stress is calculated by taking the wall unit normal projection as follows:
\begin{equation}
\Bar t = \Bar{\Bar\tau} \cdot \hat n
\end{equation}
Here, $\hat n$ is the unit normal vector of the artery wall.\\
The wall shear stress vector is calculated as the tangential components of the traction vector,
\begin{equation}
\Bar{\tau}_{w} = \Bar t - (\Bar t \cdot \hat n)\hat n
\end{equation}
} The magnitude of the strain rate tensor field can be calculated as $\dot \gamma = \sqrt{2\Bar{ \Bar{ \gamma}} \colon \Bar{ \Bar{ \gamma}}}$.
Wall shear stress is calculated for both patient-specific and simplified geometries (at all levels) and presented in the form of contours. The wall shear stress (WSS) contours are shown in Figure \ref{WSS} for all the time instants of the pulsatile cycle for the original and other geometric simplification levels while adopting the Apostolidis and Beris viscosity model. The features of WSS in Level 1 are particularly significant in predicting the onset and progression of disease, as they provide detailed insights and identify critical regions. 
Figure \ref{WSS} illustrates how the simulation accuracy deteriorates as the geometry is simplified. Firstly, for Level 1, with all perturbations preserved, we observe high variability in WSS at the peak instant of flow. At all times, the surface perturbations and the asymmetry generate several high (and low) stress zones of WSS. Some of these variabilities are captured in Level 2, although the asymmetry of Level 1 is lost. The high and low-stress zones appear as bands around the circumference, with an overall reduction of the variations observed in the WSS distributions of Level 1. This is most pronounced at the time instant of peak velocity. However, the WSS profile is smooth at Level 3 (tube of decreasing area), and all the variability is lost. Thus, Level 2 might still provide satisfactory predictions of WSS, which is entirely lost in Level 3. \textcolor{blue}{The variability in these hemodynamic parameters, in terms of maximum and minimum values, is presented in Table \ref{WSS_table}. The calculated wall shear stress falls within a realistic range. Note that the time-averaged wall shear stress values fall within the limits of the study by Moore Jr. et al.\cite{moore1994fluid}}. For simplicity, we present the data for all three viscosity models at only two time instants of the cycle: \textit{t} = 0 s and \textit{t} = 0.35 s (peak). The table further highlights the discussion about the accuracy of Levels 2 and 3 (relative to Level 1) presented above. The table shows that while the values in Level 2 retain some similarity, the values for Level 3 are significantly different in most cases. These trends broadly fit the discussion presented above.

\begin{table}[htbp]
\centering
  \caption{Minimum and maximum wall shear stress magnitude values (in Pa) recorded at the start (\textit{t} = 0 s) and at the peak (\textit{t} = 0.35 s) of the cardiac cycle for the simplified geometrical levels and tabulated for the selected viscosity models.}

\renewcommand{\arraystretch}{1.5} 

\begin{tabular}{|c|c|c|c|c|c|c|c|}
\hline
\multirow{2}{*}{time} & \multicolumn{1}{c|}{Geometry $\rightarrow$} & \multicolumn{2}{c|}{Level 1} & \multicolumn{2}{c|}{Level 2} & \multicolumn{2}{c|}{Level 3} \\
\cline{2-8}
& \multicolumn{1}{c|}{Viscosity models $\downarrow$} & min & max & min & max & min & max \\

\hline
\multirow{3}{*}{\textit{t} = 0 s} 
& Apostolidis \& Beris & 0.170 & 0.647 & 0.209 & 0.686 & 0.230 & 0.527 \\
& Carreau-Yasuda       & 0.014 & 0.398 & 0.037 & 0.371 & 0.067 & 0.229 \\
& Newtonian            & 0.011 & 0.401 & 0.394 & 1.385 & 0.062 & 0.223 \\
\hline
\multirow{3}{*}{\textit{t} = 0.35 s} 
& Apostolidis \& Beris & 0.029 & 5.930 & 0.004 & 3.960 & 0.378 & 1.444 \\
& Carreau-Yasuda       & 0.006 & 2.642 & 0.000 & 2.807 & 0.100 & 0.764 \\
& Newtonian            & 0.004 & 2.642 & 0.001 & 2.949 & 0.094 & 0.756 \\
\hline
\end{tabular}

    \label{WSS_table}
\end{table}

\subsubsection{\large Correlations}
The previous discussion on retention of accuracy is qualitative. To enable the quantification of the impact of simplified geometric levels on accuracy, we introduce a correlation metric here. These correlations can be used to compare the accuracy levels of any hemodynamic parameter across the descending aorta. Here, we show the definition of the correlation for the WSS, but it can also be extended to pressure. The calculations use the deviations of any parameter from its mean at any location, as follows:
\begin{equation}
\theta =  \text{WSS}_{c} - \overline{\text{WSS$_{c}$}};\ \text{for wall shear stress.}
\end{equation}
Here, $\overline{\text{WSS$_{c}$}}$ is the average over the length of the artery, circumferential average wall shear stress (\text{WSS}$_{c}$) at any location, and $\theta$ is the deviation parameter. The deviation parameter $\theta$ is calculated for the simplified levels of the descending aorta, i.e., $\theta_{1}$,$\theta_{2}$, and $\theta_{3}$ for Level 1, Level 2, and Level 3, respectively. Keeping Level 1 as a reference, the correlation $(C)$ between the different geometry levels is defined as:

\begin{equation}
    C_{11} = \frac{\sum_{z} \theta_{1}\theta_{1}}{\sum_{z} \theta_{1}\theta_{1}} = 1;
    C_{12} = \frac{\sum_{z} \theta_{1}\theta_{2}}{\sum_{z} \theta_{1}\theta_{1}};
    C_{13} = \frac{\sum_{z} \theta_{1}\theta_{3}}{\sum_{z} \theta_{1}\theta_{1}};
    C_{23} = \frac{\sum_{z} \theta_{2}\theta_{3}}{\sum_{z} \theta_{2}\theta_{2}};
\end{equation}
where $C_{12}$, $C_{13}$, and $C_{23}$ are the correlation parameters between levels 1 and 2, levels 1 and 3, and levels 2 and 3, respectively. By definition, $C_{11}$ should be unity and is used in the plots for validation. Thus, correlation parameters provide a sense of a “hotspot” (in WSS or pressure) being observed at a similar location when two different geometry levels are used. If a simplified geometry can capture all hotspots of level 1 at their respective locations, the value of the correlation parameter would be close to unity.
\par
\begin{figure}[H]
    \centering
    \includegraphics[width=1.02\linewidth]{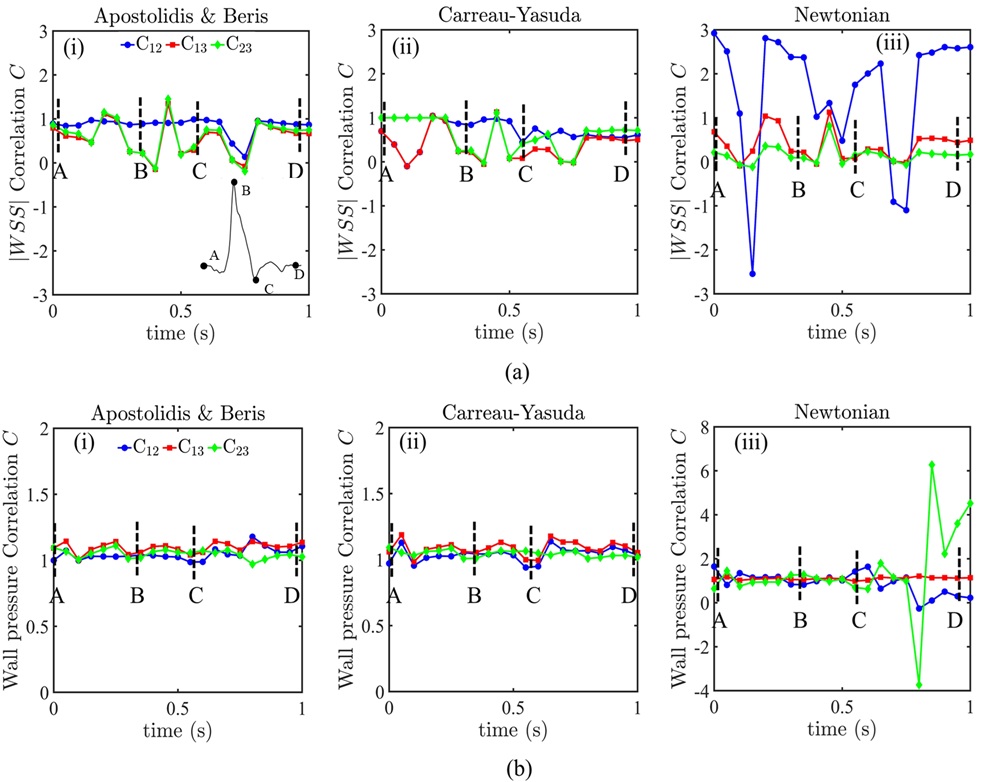}
  \caption{Correlations of (a) wall shear stress magnitude and (b) wall pressure are plotted with time for the selected viscosity model, i.e, (i) Apostilidis and Beris, (ii) Carreau-Yasuda, and (iii) Newtonian. The correlation parameter values are highlighted at phases of the velocity waveform given at \textit{t} = 0 s (A), \textit{t} = 0.35 s (B), \textit{t} = 0.55 s (C), and \textit{t} = 0.9 s (D).}
    \label{Correlations}
\end{figure}

 Here, we evaluate correlation levels for WSS and pressure. These correlations are computed at specific time instants of the waveform. The correlation coefficient, \( C \), is evaluated for the wall shear stress (WSS) magnitude and wall pressure over the entire cardiac cycle for the selected viscosity models, and the results are presented in Figure \ref{Correlations}. The correlation trends for both WSS (Figure \ref{Correlations}(a)) and wall pressure (Figure \ref{Correlations}(b)) are examined. The WSS values of $C_{12}$ remain close to unity for the non-Newtonian models but show large fluctuations for the Newtonian model. In contrast, $C_{13}$ and $C_{23}$ remain further away from unity. This quantitatively supports our preceding discussion. For WSS, Level 2 predictions remain highly correlated with Level 1 (and thus retain accuracy), whereas much of this accuracy is lost in Level 3. However, the Newtonian model is not well-behaved for these simulations, exhibiting large fluctuations in correlation values across models. Particularly, the cross-correlations $C_{12}$ and $C_{13}$ are large during phases C and D for the Newtonian simulation, suggesting the role of the flow structures that appear in Level 1. This suggests that non-Newtonian models of blood may be more suitable for simulations in geometries with a high degree of irregularity. Conversely,  accuracy loss is observed to be mild in pressure values.

\subsubsection{\large Time-averaged wall shear stress (TAWSS)}
\begin{figure}[H]
    \centering
    \includegraphics[width=0.8\linewidth]{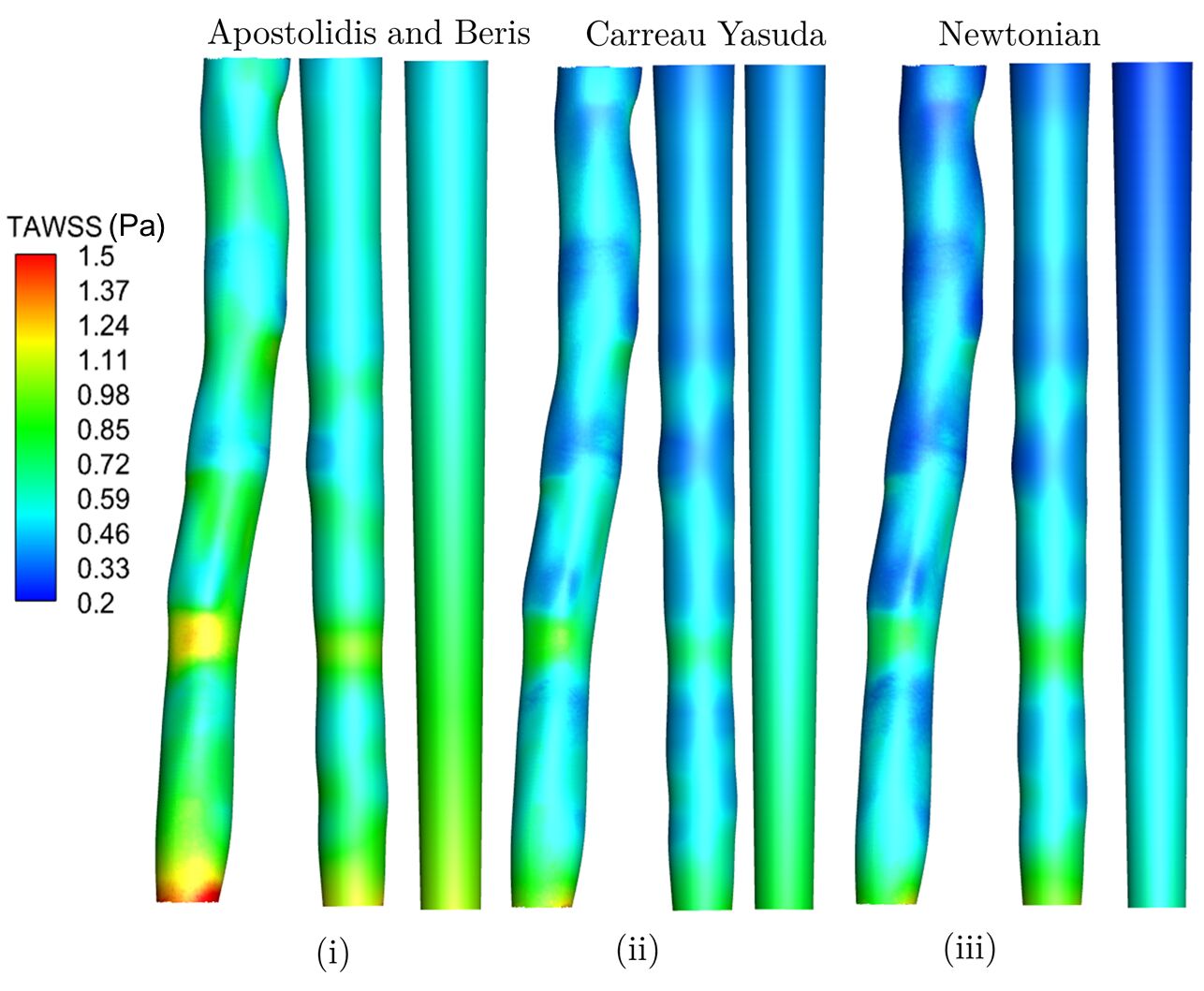}
  \caption{Time average wall shear stress (TAWSS) shown for the three viscosity models for the Levels 1,2, and 3 geometry of the descending aorta. (a) Apostolidis and Beris, (b) Carreau-Yasuda and (c) Newtonian viscosity model.}
    \label{TAWSS}
\end{figure}

Along with WSS, time-averaged wall shear stress (TAWSS) is a primary risk parameter to monitor the development and progression of atherosclerosis, which can further lead to various types of cardiovascular diseases. Both low- and high-wall shear stress promote pro-inflammatory and anti-inflammatory responses in endothelial cells [31]. An average WSS over the cardiac cycle is the time-averaged wall shear stress (TAWSS), which further helps identify the high and low WSS zones. Mathematically, TAWSS is defined by the following relation:
\begin{equation}
\text{TAWSS} = \frac{1}{T}\int^{T}_{0} |\Bar\tau_{w}| dt
\end{equation}
Here, $T$ is the time period of the cardiac cycle.
The time-averaged wall shear stress (TAWSS) calculated for the fourth cycle of the simulations is shown in Figure \ref{TAWSS} for the three levels of the geometry of the descending aorta and the selected viscosity models.
In Figures \ref{TAWSS} (i),(ii), and (iii), TAWSS contours are plotted for Apostolidis and Beris, Carreau Yasuda, and Newtonian viscosity models, respectively. The trends are qualitatively similar to those discussed earlier in WSS. In Level 1, the perturbations and asymmetry of the geometry generate a significant number of zones for high and low TAWSS. Some of these zones are retained in Level 2 with less asymmetry and a reduction in the shear stress magnitudes. A gradual variation of TAWSS is observed in Level 3, characterized by a smooth geometry. The trends are similar across the three viscosity models. However, the perturbations in TAWSS are slightly different from those in the Apostolidis and Beris model, whereas the profiles are identical for the Newtonian and Carreau-Yasuda models. The values are somewhat higher with the Apostolidis and Beris model in Level 1. Thus, again, concerning spots of extreme stress, Level 2 captures the essence partially, while variations in Level 3 are averaged out.

\subsubsection{\large Oscillatory shear index (OSI)}
In pulsatile flow, the shear stress near the artery wall fluctuates, making the endothelial cell deliver a pro-inflammatory response. The Oscillatory Shear Index (OSI) introduced by Ku et al. \cite{ku1983pulsatile} provides a measure of these fluctuations. The OSI can be calculated as follows:
\begin{equation}
\text{OSI} = \frac{1}{2}\left(1-\frac{|\int^{T}_{0} \Bar\tau_{w} dt|}{\int^{T}_{0} |\Bar\tau_{w}| dt}\right)
\end{equation}
\noindent OSI varies between 0 and 0.5, the latter representing a purely oscillatory response. The high-value region exhibits a pronounced time-periodic oscillation in the wall shear stress. 
\begin{figure}[H]
    \centering
    \includegraphics[width=0.8\linewidth]{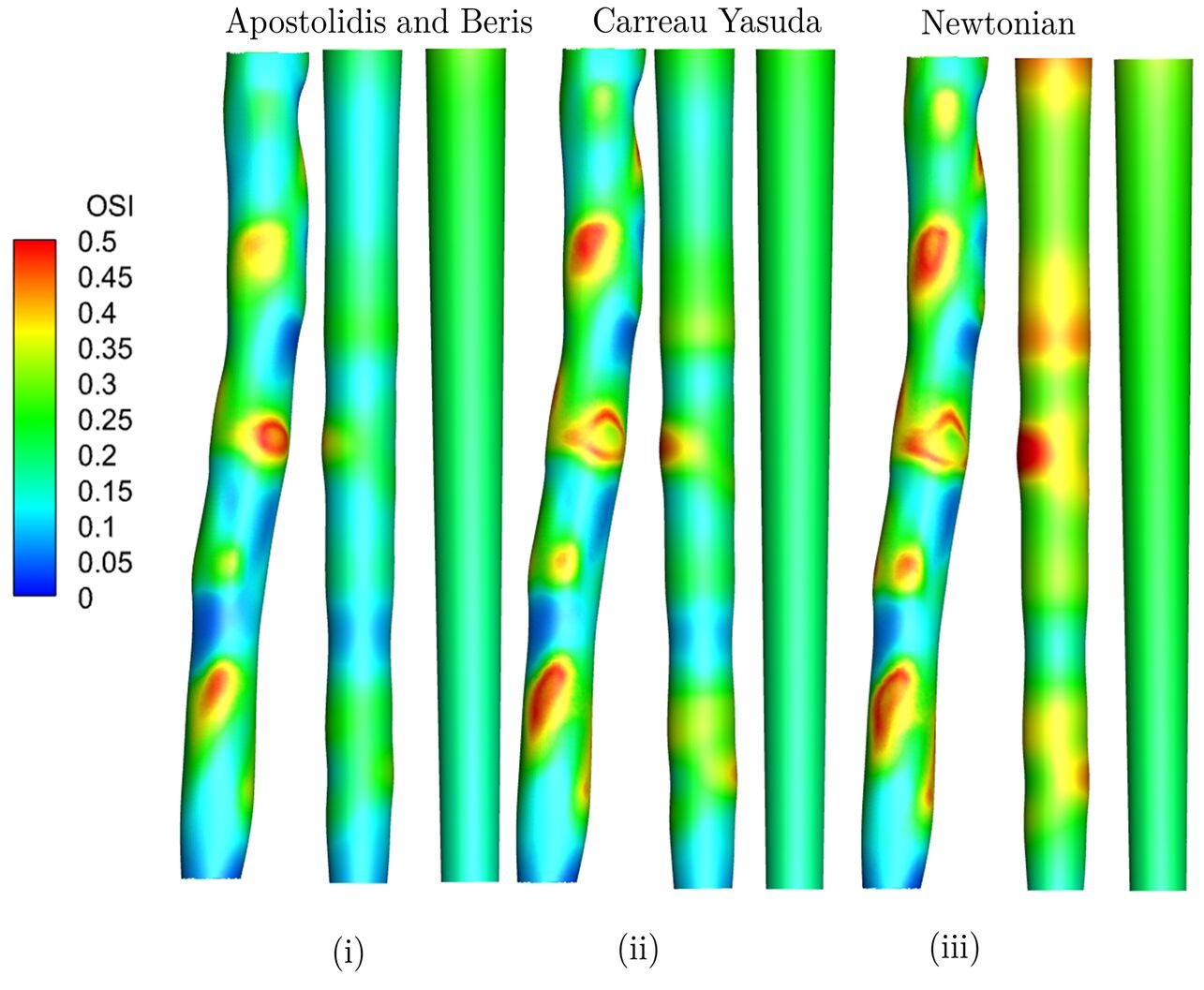}
  \caption{Oscillatory shear index (OSI) shown for the three viscosity models for Levels 1,2, and 3 geometry of the descending aorta. (a) Apostolidis and Beris, (b) Carreau-Yasuda and (c) Newtonian viscosity model.}
    \label{OSI}
\end{figure} 
Figure \ref{OSI} plots Oscillatory Shear Index contours for the selected viscosity models. OSI contours depict the zone of shear oscillation and its intensity. In Figure \ref{OSI}, the OSI distributed zones in the Level 1 artery can be seen with hotspots of high and low OSI values, signifying the nature of shear stress oscillations. The region with high OSI values is more prone to a pro-inflammatory response by the endothelial cells, whereas the low OSI value indicates unidirectional shear stress; therefore, an anti-inflammatory response is obtained for the cell. As the complexity of the descending aorta gradually decreases, the OSI zones are also averaged uniformly, as the geometrical features are refined in the subsequent levels of simplification. Lastly, a uniform OSI distribution is observed in Level 3, as no geometrical disturbances are present.
The asymmetry of Level 1 is lost in both Levels 2 and 3 due to the simplification of the geometric structure.\par
\begin{figure}[H]
    \centering
    \includegraphics[width=1\linewidth]{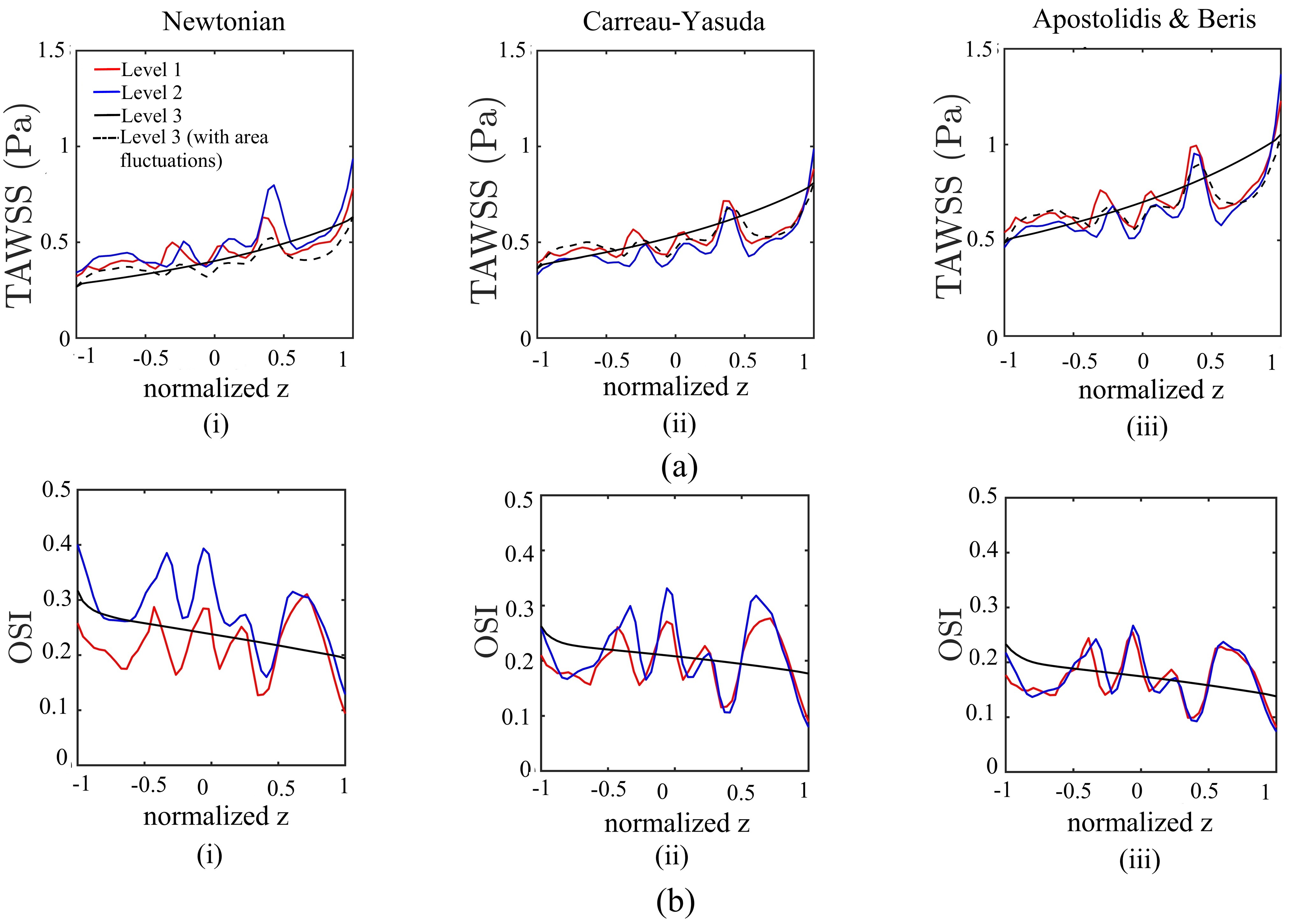}
  \caption{Comparison of TAWSS and OSI for (i) Newtonian, (ii) Carreau-Yasuda, and (iii) Apostolidis and Beris models. (a) TAWSS: Time-Averaged Wall Shear Stress. (b) OSI: Oscillatory Shear Index for the geometric simplification levels 1-3 of the descending aorta. The results of Level 1 shown here are averaged over the circumference for each axial location.}
    \label{time_cum_circ_TAWSS_OSI}
\end{figure}

 A fair comparison of the time-averaged hemodynamic indicators across geometric levels is shown in Fig. \ref{time_cum_circ_TAWSS_OSI} for TAWSS (Fig. \ref {time_cum_circ_TAWSS_OSI} (a)) and OSI (Fig.\ref {time_cum_circ_TAWSS_OSI} (b)) averaged over the circumference at each axial location. As might be expected, the predictions in Level 3 are a monotonic function, while predictions in Levels 1 and 2 show oscillations due to axial variations in the geometry. For all data shown, the Level 3 predictions represent an average behavior of the oscillatory predictions obtained for Levels 1 and 2. The usefulness of Level 2 is readily apparent in these results, as it mostly closely follows the predictions of Level 1. In the present discussion, the Level 1 predictions are averaged over the circumference, and the Level 2 geometry is built by averaging diameters at various axial locations. The consistency of Levels 1 and 2 is more pronounced for the non-Newtonian models, revealing some differences for the Newtonian fluid, particularly for OSI. These results support the proposition that a somewhat smoother geometry (like Level 2) can capture fluctuations in hemodynamic parameters, albeit at the expense of local asymmetry. As discussed later in Section 3.5, this presents a valid ground for a speed-accuracy trade-off since convergence in the smoother geometries is faster and yet captures the essence of the predictions of Level 1. However, only Level 1 (patient-specific geometry) can be used without any further simplifications in applications where the details of local asymmetry are required.

For TAWSS, Level 2 can provide predictions that are close to those of the circumferentially averaged Level 1 data. However, Level 3 is only able to display a monotonic function that represents an overall average trend. It lacks the oscillation and fluctuations caused by area fluctuations, characteristic of Levels 1 and 2. It is possible to ask, from the prediction of TAWSS of Level 3 and knowing the area changes, is it possible to recover the prediction of Levels 1 and 2? The following approach is adopted here. Shear stress is given by 
$$\text{TAWSS} \sim \mu_{app}\frac{\partial u}{\partial z}$$ 
for flow in a duct. In terms of the order of magnitudes, the expression can be written as
$$\text{TAWSS} \sim \mu_{app}\frac{U}{D}\frac{\partial u*}{\partial z*}$$
Here, symbols $U$ and $D$ are the characteristic velocity and diameter at any location within the duct. The average velocity at any location can be estimated by knowing the volumetric flow rate and local cross-sectional area. Since we know the total volumetric flow rate and area at any location, $U$ and $D$ can be estimated for local perturbations while $\frac{u*}{z*}$ can be assumed to be unaffected. At a high shear rate, $\mu_{app}$ can be assumed to be a limiting value (Figure \ref{Viscosity}). Thus, TAWSS can be estimated locally from the characteristic velocity and diameter at every cross-section. These estimates are shown in Figure \ref{time_cum_circ_TAWSS_OSI}(a) for all viscosity models by black dashed lines.\\
These results clearly indicate the success of the estimated TAWSS across all models, with non-Newtonian models performing slightly better. Thus, knowing the results of the simulation for Level 3 and the prescribed area perturbations, the realistic local TAWSS in a patient-specific geometry can be readily estimated.

\subsection{$Q$ criterion}
\begin{figure}[H]
    \centering
    \begin{subfigure}[b]{\textwidth} 
        \centering
        \includegraphics[width=\textwidth]{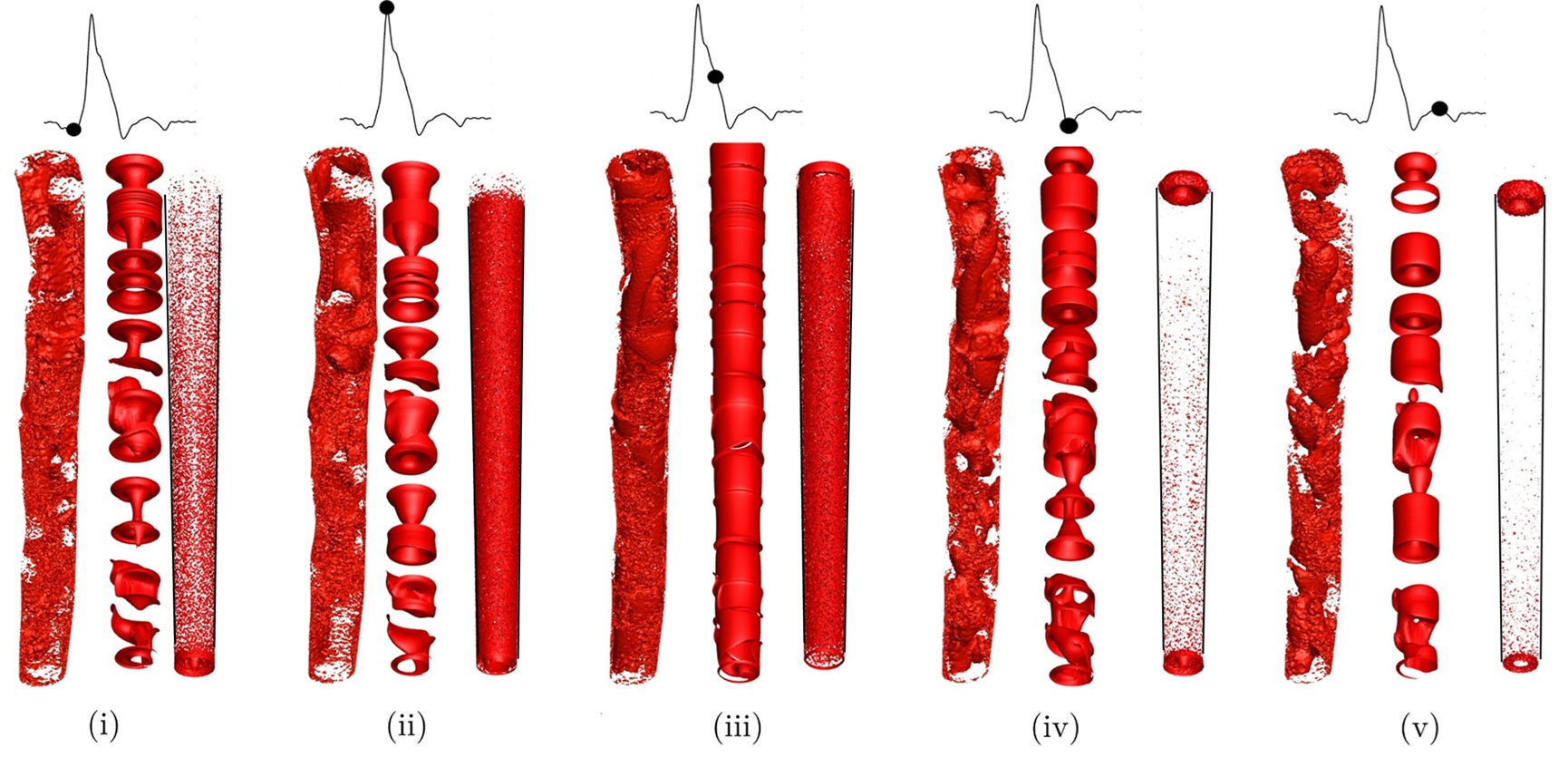}
        \caption{}
        \label{QCpos}
    \end{subfigure}
    \hfill
    \begin{subfigure}[b]{\textwidth}
        \centering
        \includegraphics[width=\textwidth]{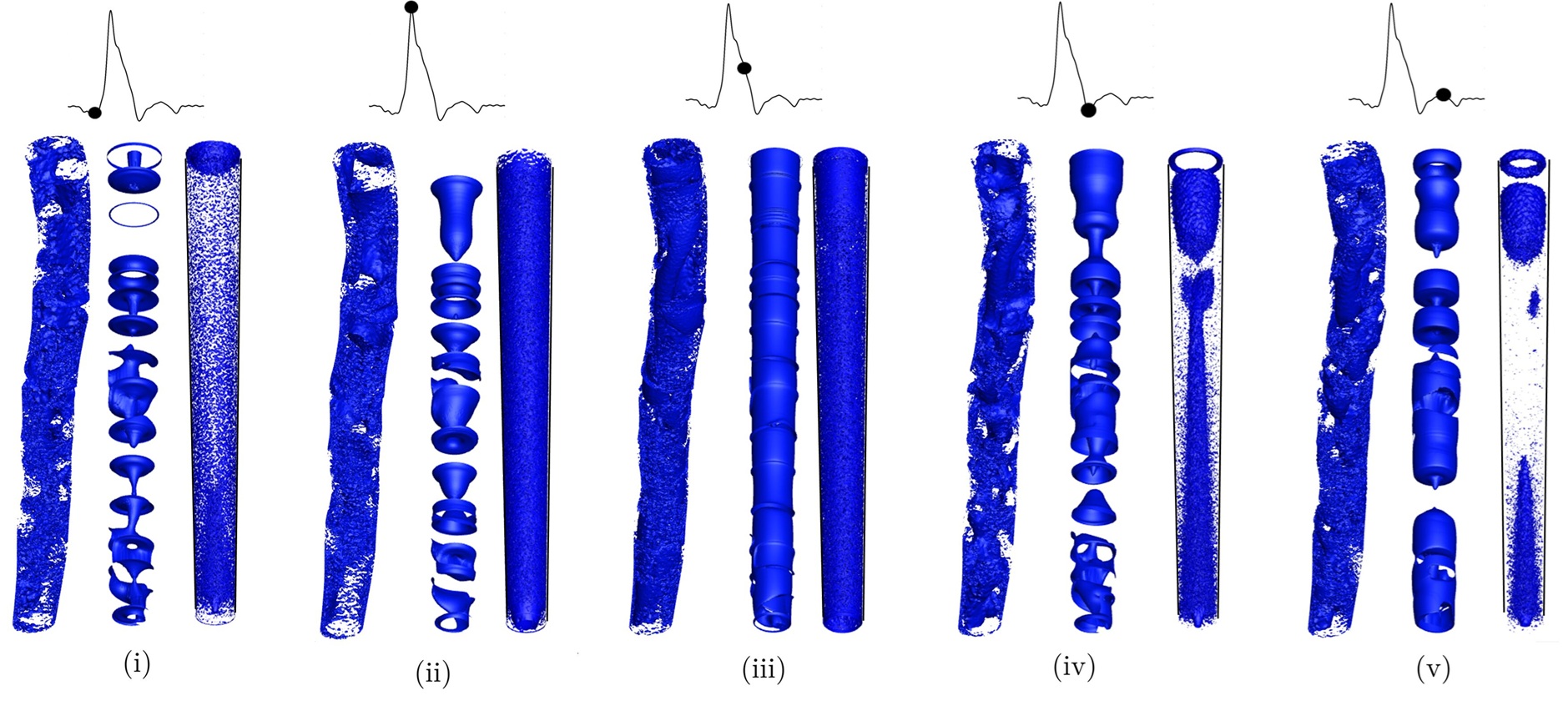}
        \caption{}
        \label{QCneg}
    \end{subfigure}
    \caption{Iso-surfaces of the $Q$ criterion are shown for dimensional values of (a) 0.1 and (b) - 0.1 values for the three geometries simulated with the Apostolidis and Beris viscosity model at the selected time instants of the vascular waveform. Within time instants of (i) - (v), the three levels are arranged from left to right.}
    \label{QC}
\end{figure}

Another critical parameter for studying strain rate and rotational characteristics is the $Q$ criterion. It is based on the velocity gradient tensor, which is given as the summation of a symmetric and anti-symmetric part as follows:
$$ \nabla \textbf{u} = \frac{ \nabla \textbf{u} +  \nabla \textbf{u}^{T}}{2} + \frac{ \nabla \textbf{u} - \nabla \textbf{u}^{T}}{2}$$ The first term is the symmetric part and represents the rate of strain tensor (${\bar{\bar\gamma}}$). The second term represents the anti-symmetric part of the vorticity tensor (${\bar{\bar\Omega}}$).
The $Q$ criterion is then calculated as:
\begin{equation}
Q = \frac{||\bar{\bar\Omega}|| - ||\bar{\bar\gamma}||}{2}
\end{equation}
A positive $Q$ value indicates a region dominated by rotational flow. Similarly, a negative value denotes a zone where rotational flows are not dominant. In Figure \ref{QCpos}, an iso-surface of the $Q$-criterion is shown for $Q$ = 0.1. Similarly, in Figure \ref{QCneg}, it is shown for $Q$ = -0.1. The simplification of the artery leads to a uniform flow distribution. Strong vortex structures are created in Level 1 and progressively disintegrate in Level 2 and Level 3. 
\renewcommand{\thesubfigure}{\alph{subfigure}}
In the deceleration phase, the $Q$ function swells out, filling the tube volume. During diastolic flow, the $Q$ contour created in Level 3 occupies a smaller volume within the tube. Overall, the vortex structure weakens considerably as the geometry is simplified. A similar trend is also seen for the contours of $Q$ = -0.1, as shown in Figure \ref{QCneg}.  The nature of the distribution of the positive values of the $Q$ function is associated with wall shear stress. In contrast, negative $Q$ values indicate that strain dominates over rotation, signifying zones of intense deformation rather than vortex formation.
\subsection {Speed Accuracy trade-off}
In the preceding sections, we have shown details of the accuracy aspects arising from various simplification levels. As discussed, Level 2 mostly captures the local effects for all parameters, but the local asymmetry visible in the original Level 1 is lost. Level 3 is unable to capture any of the regional effects and only shows a “representative average” of the results obtained in Level 1. However, for the wall shear stress, simulation data of Level 3 can be reconstructed using scaling principles, which agree well with the detailed calculations of Level 1. Thus, each level has its own utility and can provide limited but valuable information about the flow field. In this section, we discuss the reduction in computational time achieved through geometric simplification. In Level 2 geometry, we used 25 individual slices over the length of the artery, each slice being a cylindrical patch. To analyze the effect of this number, we reconstructed the entire artery by extracting the cross-sectional area of the Level 1 geometry along its length. We then developed multiple equivalent geometries consisting of 25, 20, 15, 10, and 5 slices, apart from Level 3, which has no slices. Using a grid-independence test, we analysed all reconstructed geometries with Newtonian blood viscosity under steady flow conditions, employing a uniform inlet velocity. 

\begin{figure}[H]
    \centering
    \includegraphics[width=\linewidth]{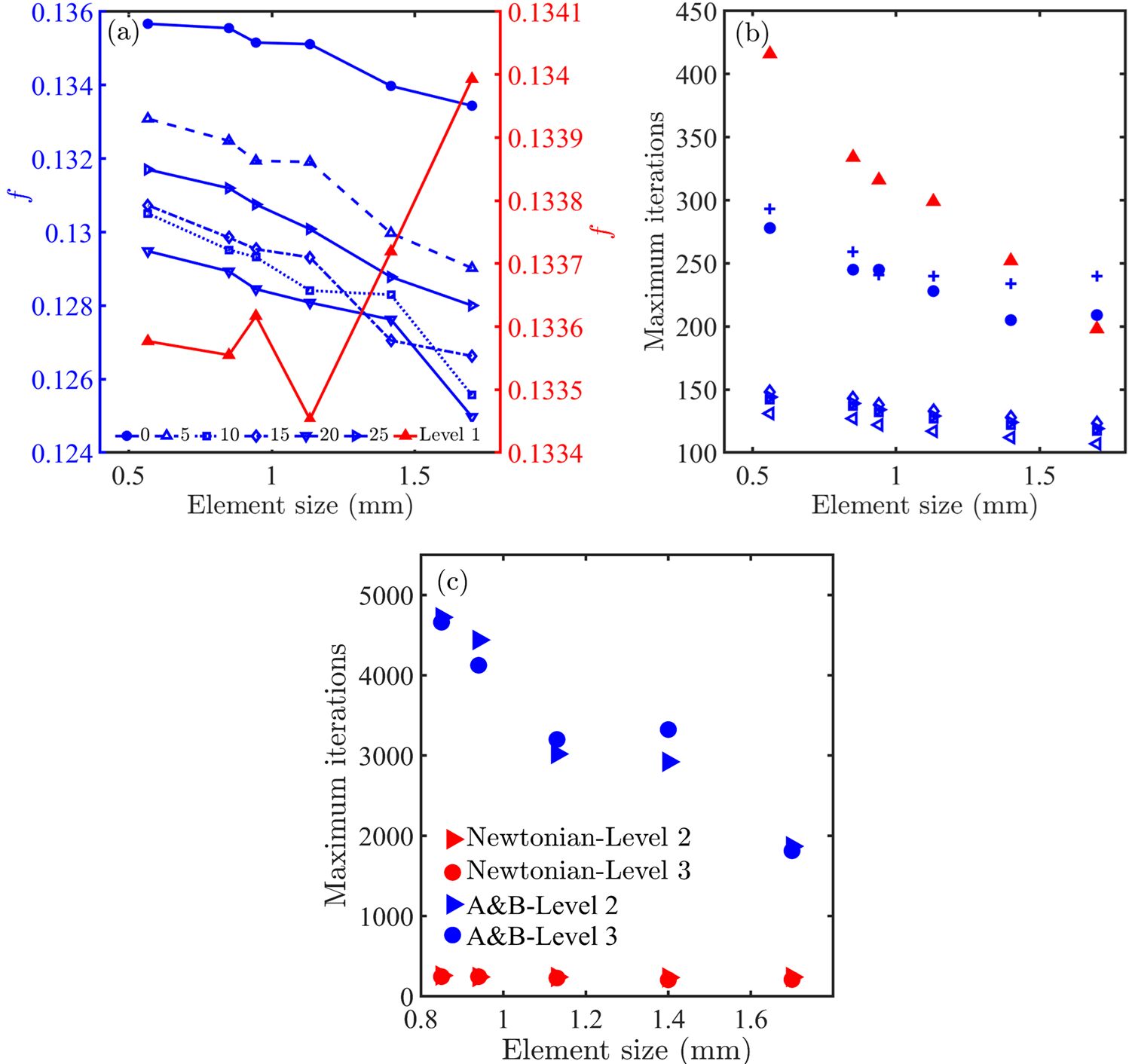}
    \caption{ Speed-accuracy analysis is carried out by performing (a) grid independence test for simplified geometries up to 25 slices and comparing their respective friction factors. (b) The maximum number of iterations required to reach the tolerance of 1e-5 in velocity is plotted as a function of the element size. (c) The maximum number of iterations is plotted for the Apostolidis and Beris (A$\&$B) and Newtonian viscosity models for Level 2 (with 25 slices) and Level 3 simplification.}
    \label{SAT}
\end{figure}

In Figure \ref{SAT}(a), we show the friction factor for all reconstructed geometries, including the Level 1 geometry, as a function of the mesh element size. For Level 1 geometry, the converged mesh size appears to be around 0.6. Similarly, the friction factor for all other simplified geometries is plotted, with convergence occurring at about an element size of 0.8. The results indicate that the simplified geometries require coarser meshes to achieve grid convergence compared to the Level 1 geometry. This finding aligns with expectations, as the simplified geometries are symmetrical along their axis and less complex, making convergence attainable on coarser meshes. Consequently, the simplified arteries are expected to yield faster simulations than Level 1. Apart from the asymmetry, most of the local flow aspects are resolved by the Level 2 geometry. Thus, we can perform simulations to a similar level of accuracy using a coarser mesh, thereby increasing the computational efficiency of the simulation.
Further, to show the speed increase with geometric simplification, we show the maximum number of iterations required to reach a prescribed convergence criterion as a function of the element size, in Figure \ref{SAT}(b). The data reveal that simplified geometries with 5 to 20 slices require approximately 150 iterations. In contrast, the simplified artery with 25 slices and the geometry with zero slices need around 300 iterations for the smallest element size and 200 iterations for the largest element size. For Level 1 geometry, the number of iterations increases as the element size is decreased, which is consistent with expectations for unstructured meshes over complex three-dimensional geometries. Thus, there is a clear trend of improvement in terms of the number of iterations (and, consequently, the speed of computation) as we move towards simplified geometric representations. The difference is as large as four times between Level 1 and Level 2 (with 25 slices) at their optimal element sizes. 

In Figure \ref{SAT}(c), the Apostolidis and Beris viscosity model is compared with the Newtonian viscosity model with respect to the maximum number of iterations as a function of element size for Level 2 and Level 3 geometries. A significant difference is observed when the viscosity model becomes non-Newtonian. For the converged mesh, nearly 5,000 iterations are required for the non-Newtonian viscosity model compared to the Newtonian model. Overall, for meshes with acceptable grid quality, geometric simplification reduces the number of iterations required, while the nonlinearity of the viscosity model increases the number of iterations; grid refinement, on the other hand, increases the number of iterations.
\section{Discussion}
The study provides two broad observations. First, geometric simplifications can help generate fast but approximate insights, primarily when time constraints exist in clinical decision-making. The approximate solutions can often be improved using suitable scale analysis. However, for a detailed analysis, retaining complete geometrical fidelity is essential. Secondly, for larger arteries, the Newtonian model can provide qualitatively similar predictions faster than the non-Newtonian ones in terms of CPU time. The findings also emphasize the importance of combining accurate hemodynamic indicators, such as time-averaged wall shear stress and the $Q$-function, with newly developed blood viscosity models for better prediction of blood-tissue interaction. These insights could aid in designing personalised treatments, evaluating surgical interventions, and improving diagnostic accuracy in detecting the onset of cardiovascular diseases.
\section {Conclusions}
A detailed study of a patient-specific geometry of the descending aorta, with two simplified representations and three viscosity models, is conducted. The geometry is extracted from CT scan data shared by a hospital and reconstructed using image processing tools. Realistic time-dependent boundary conditions are applied at the inlet and outlet, respectively, in terms of pulsatile velocity (mean Re = 313 and peak Re = 1687) and outlet pressure. For the geometrical approximation, the original artery (termed Level 1) is further simplified into two more levels, denoted as Level 2 and Level 3. Level 2 is based on the cross-sections of Level 1, which is divided into 25 sections, with each part having a representative average diameter. Thus, Level 2 captures some of the undulations and area fluctuations of Level 1 but loses the local asymmetry of the original. Level 3 represents a smooth duct based on the inlet and outlet cross-sectional area of the actual artery. The 3D simulations are performed using three blood rheology models: Newtonian, Carreau-Yasuda, and Apostolidis and Beris. Various aspects of blood flow for all the levels of the descending aorta geometry (with selected blood viscosity models) are studied. The following conclusions are drawn in this work.
\begin{enumerate}
\item
As the geometrical complexity is averaged out, the overall flow field becomes uniform, with a reduction in the strength of the recirculation zones near wall perturbations. For Level 1 geometry, the flow field shows a significant number of recirculation zones near the walls. Some of these zones are also visible in Level 2, albeit with reduced intensity. No such zone of reversed flow is visible in Level 3, where the flow field is practically unidirectional. 
\item
The simplification of the geometry affects the strain rate distribution, which in turn influences the wall shear stress (WSS). Simulations indicate that Level 2 geometry can reasonably approximate the WSS distribution, capturing major patterns and magnitudes while missing the local asymmetry in the stress profiles. However, Level 3 shows only a smooth variation of the WSS without showing undulations.

\item 
The correlation analysis indicates that Level 2 prediction of WSS maintains a strong agreement with Level 1 results, thereby preserving greater accuracy when employing non-Newtonian models. In contrast, this accuracy significantly diminishes in Levels 2 and 3  when using Newtonian viscosity. Pressure remains well-correlated, except during phases C and D of the diastolic cycle, for the Newtonian model.

\item

The variations in hemodynamic indicators, such as TAWSS and OSI, further bolster these earlier assertions. Level 2 predictions are particularly close to those obtained in Level 1, especially when both sets of data are averaged over the circumference. However, Level 3 predictions are without any fluctuations and are, at best, a representative average. 
\item
A simple estimate of the actual WSS in the original artery can be obtained by knowing the prediction from Level 3, scaled by the area fluctuation. The estimates thus retrieved are close to the ones obtained with Level 1.
\item  
The speed test revealed that computations with Levels 2 and 3 were approximately four times faster than those with Level 1. The Newtonian model requires approximately 10 times fewer iterations than the Apostolidis and Beris model to achieve a comparable convergence level and grid refinement.

\end{enumerate}
\textbf{Acknowledgment}\\
Authors acknowledge the National Supercomputing Mission (NSM) for providing computing resources of 'PARAM Sanganak' at IIT Kanpur, which is implemented by C-DAC and supported by the Ministry of Electronics and Information Technology (MeitY) and Department of Science and Technology (DST), Government of India. Additionally, we would like to thank the Computer Centre (www.iitk.ac.in/cc) at IIT Kanpur for providing the resources to carry out the reported work. The authors gratefully acknowledge the kind support provided by the DST NSM project with DST/NSM/R\&D\_HPC\_Applications/2021/18 sanction number .\\
\textbf{Author Declaration}\\
The authors declare that they have no conflicts of interest.

\bibliographystyle{apsrev4-2}
\bibliography{references}

\end{document}